\documentclass[5p]{elsarticle}
\usepackage{graphicx}
\usepackage{epstopdf,subfigure}
\usepackage{amssymb}\usepackage{url}
\usepackage{natbib}
\usepackage{url}
\bibpunct{(}{)}{;}{a}{}{,}

\journal{Icarus}

\begin{document}

\begin{frontmatter}

%\thesaurus{10(03.07.1, 13.03.2, 18.05.1)}

\title{Elliptical instability in hot Jupiter systems}

\author[eth,irphe]{David~C\'ebron\corref{cor1}}
\ead{david.cebron@erdw.ethz.ch}
%\fntext[label1]{}
\cortext[cor1]{Corresponding author.}
\author[irphe]{Michael~Le~Bars}
\author[irphe]{Patrice~Le~Gal}
\author[lam]{Claire~Moutou}
\author[LMD]{Jeremy~Leconte}
\author[irphe]{Alban~Sauret}

%\offprints{J. Crovisier;\\
%\email{Jacques.Crovisier@obspm.fr}}

\address[eth]{Institut f\"ur Geophysik, Sonneggstrasse 5, ETH Z\"urich, Z\"urich, CH-8092, Switzerland.}
\address[irphe]{Institut de Recherche sur les Ph\'enom\`enes Hors Equilibre, UMR 7342, CNRS and Aix-Marseille Universit\'e, 49 rue F. Joliot-Curie, B.P. 146, 13384 Marseille cedex 13, France.}
\address[LMD]{Laboratoire de M\'et\'eorologie Dynamique, Institut Pierre Simon Laplace, Paris, France}
\address[lam]{Observatoire Astronomique de Marseille-Provence, Laboratoire d'Astrophysique de Marseille, 38 rue F. Joliot-Curie, 13388 Marseille cedex 13, France.}

% \date{Received  / Accepted}

\date{Draft \today}

%\titlerunning{103P/Hartley 2 observed at Nan\c{c}ay}

%\authorrunning{Crovisier et al.}

\begin{abstract}

{Several studies have already considered the influence of tides {on} the evolution of systems composed of a star and a close-in companion to tentatively explain different observations such as the spin-up of some stars with hot Jupiters, the radius anomaly of short orbital period planets and the synchronization or quasi-synchronization of the stellar spin in some extreme cases. However, the nature of the mechanism responsible for the tidal dissipation in such systems remains uncertain. In this paper, we claim that the so-called elliptical instability may play a major role in these systems, explaining some systematic features present in the observations. This hydrodynamic instability, arising in rotating flows with elliptical streamlines, is suspected to be present in both planet and star of such systems, which are elliptically deformed by tides.}
   {The presence and the influence of the elliptical instability in gaseous bodies, such as stars or hot Jupiters, are most of the time neglected.}
   {In this paper, using numerical simulations and theoretical arguments, we {consider} several features associated to the elliptical instability in hot-Jupiter systems. In particular, the use of ad hoc boundary conditions makes it possible to estimate the amplitude of the elliptical instability in gaseous bodies. We also consider the influence of compressibility on the elliptical instability, and compare the results to the incompressible case.}
   {We {demonstrate} the ability {for} the elliptical instability to grow in the presence of differential rotation, with a possible synchronized latitude, provided that the tidal deformation and/or the rotation rate of the fluid are large enough. Moreover, the amplitude of the instability for a centrally-condensed mass of fluid is of the same order of magnitude as for an incompressible fluid for a given distance to the threshold of the instability. {Finally, we show that the {assumption} of the elliptical instability being the main tidal dissipation process in eccentric inflated hot Jupiters and misaligned stars is consistent with current data.}}

\end{abstract}
\begin{keyword}

Tides; Stars; hot Jupiters; Elliptical instability; Radius anomaly; Spin-orbit misalignment

\end{keyword}

\end{frontmatter}

\section{Introduction}
%%%%%%%%%%%%%%%%%%%%%%

\subsection{Tides in extrasolar planetary systems}

The search for planetary systems shows that a substantial fraction of the observed stars {hosts} extrasolar planets. Methods of detection, such as the radial-velocity method or the transit method, are most sensitive to large planets on close orbits. Consequently, many known extrasolar planets have a mass comparable to that of Jupiter and orbit very close to their host stars \cite[][]{howard2010occurrence,mayor2011exoplanets}. This population of planets, the so-called hot Jupiters, currently includes about $25\ \%$ of all known planets (see \texttt{exoplanets.org, exoplanets.eu}). The increasing amount of data on these systems leads to the possibility to test more precisely the theoretical models of interaction between celestial bodies, such as tidal or magnetic mutual influences \cite[e.g.][]{cuntz2000stellar,gu2009thermal}. Indeed, the permanent and systematic presence of tides in binary systems leads to clear observational evidence of their crucial roles \cite[see the recent review of][]{mazeh2007observational}. For instance, the ellipsoidal effect or the apsidal motion (precession of the line of apsides of a non-circular orbit, due to the mutual tidal distortion in a binary system) are clear signatures of the tidal deformations, whereas the tidally induced dissipation and angular momentum exchanges control the orbital evolution of the system, driving it toward a synchronized state on a circular orbit  \cite[excluding possible thermal tides, which can drive a planet away from synchronism: e.g.][]{gold1969atmospheric}.

The role of tides has been recently investigated for systems composed of a star and a close-in companion to tentatively explain, for instance, the spin-up of stars with hot Jupiters \cite[][]{pont2008empirical,lanza2009hot}, the radius anomaly of short orbital period planets \cite[][]{leconte2010tidal,leconte2010radius}, and the synchronisation or quasi-synchronisation of the stellar spin \cite[][]{aigrain2008transiting}. Recent observations have attracted our interest to reconsider the possible consequences of an often neglected phenomenon: the so-called elliptical instability. 

For instance, the star $\tau$ Boo has a massive planet in close orbit (4.5 $M_{Jup}$ minimum mass and 3.31-day period), and a stellar surface rotating synchronously with the planetary orbit \cite[][]{butler1997three,walker2008most}. It is possible that tides from the planet onto the star have synchronised the thin convective zone of this F7 star, since the mass ratio between the planet and the convective zone is larger than $10$ and the stellar angular momentum represents typically $60-70 \%$ of the total angular momentum. It has been possible to reconstruct the global magnetic topology of the star since 2006 and its evolution using recurrent spectropolarimetric observations. Two polarity reversals have been observed in two years \cite[][]{donati2008magnetic,fares2009magnetic}, which represent an evidence for a magnetic cycle of 800 days, much shorter than the {cycle} of the Sun (22 yr.). The role of the planetary tides on the star in this short activity cycle was questioned; a strong shear may take place at the bottom of the convective zone, triggering a more active and rapidly evolving dynamo \cite[][]{donati2008magnetic,fares2009magnetic}.

The spin-orbit misalignment (i.e. the angle between the stellar spin axis and planetary orbit normal) of one third of transiting hot Jupiters \cite[][]{winn2010hot} also questions the role of tides in such systems, since tides are responsible for alignment, circularization of the orbit and synchronisation of periods (or orbital migration, if the stellar moment of inertia of the star is too large compared to that of the planetary orbit). The usual idea of planet formation and migration within a disk was also challenged by such observations. Elliptical instabilities may cause the rotational axis of both bodies in the system to change orientation with time, with a relatively short timescale. Misaligned systems could thus show unstable rotation axes of stars, rather than tilted orbital planes of the planet. Tidal implications on the internal structure of planets were already reported \cite[][]{leconte2010tidal,leconte2010radius}. {However, the actual mechanisms responsible for this dissipation are still a matter of debate, especially considering the fact that most mechanisms predict dissipation rates that are too weak compared to the one observed in gaseous bodies. Quantifying the power generated by elliptical instabilities \cite[e.g.][for a fluid layer of a terrestrial body]{le2010tidal} is thus crucial to enhance the predictive ability of coupled orbital/planet interiors models.}

\subsection{The elliptical instability}

\begin{figure}
  \begin{center}
     \includegraphics[scale=0.45]{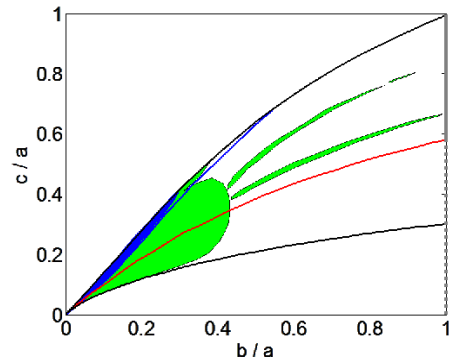}
    \caption{(Colour online) Riemann ellipsoids (for direct configurations, see \citealp{lebovitz1996new}), with $a$ (resp. $b$) the longest (resp. shortest) equatorial axis and $c$ the polar axis. The Jacobi ellipsoids (red solid line) branch off the Maclaurin spheroids (grey line, solid and dashed for stable and unstable spheroids respectively) at $c/a \approx 0.58$, and the addition of an internal uniform vorticity extends the solutions domain to the zone between the uppermost and the lowest solid black lines. For perturbations that are linear in the spatial coordinates, Riemann obtained unstable ellipsoids between the solid blue line and the black solid uppermost line, but \cite{chandrasekhar1965equilibrum,chandrasekhar1966equilibrium} showed that the correct unstable zone is the dark gray (blue) one. The light gray (green) zone corresponds to unstable ellipsoids for quadratic perturbations \cite[e.g.][]{lebovitz1996new}.}
    \label{fig:rieman}
  \end{center}
\end{figure}

The elliptical instability is a generic instability that can affect rotating fluids whose streamlines are elliptically deformed \cite[see the review by][]{kerswell2002elliptical}. A three-dimensional flow may be excited in the bulk of rotating fluid bodies when (i) the amplitude of the elliptical forcing characterized by the ellipticity of the streamlines $\beta$ is sufficiently large compared to viscous dissipation effects characterized by the Ekman number $E$ and (ii) when a difference in angular velocity
exists between the mean rotation rate of the fluid and the
elliptical distortion. In an astrophysical context, the elliptical deformation has often been related to the gravitational deformation of the fluid domain, coming from the static and periodic terms of the tidal potential. The elliptical instability has thus been suggested in tidally deformed accretion disks \cite[][]{goodman1993local,lubow1993tidal,ryu1994nonlinear,balbus1998instability,lebovitz2004magnetoelliptic} and tidally deformed stars \cite[][]{rieutord2003evolution}. But note that such an elliptical distortion can also come from a rigid boundary which has been deformed in the past and remains frozen in this deformed state, as for instance the Moon  mantle \cite[][]{lebarsNature}.  Second, this distortion can also come from a local vortex interaction in the fluid: elliptical vortices present in accretion disks could then be destabilized by the elliptical instability \cite[e.g.][]{lesur2009stability,lesur2009subcritical,lyra2010baroclinic}. Third, the ellipticity of streamlines {may} also appear spontaneously {in} rapidly rotating isolated fluid bodies. To understand this less intuitive configuration, let us consider a simple rotating homogeneous isolated fluid body. The problem of the equilibrium figure of such a body has been first solved by \cite{newton1687philosophiae} in the case of small rotation rates, and extended by \cite{maclaurin1742treatise} to arbitrary rotation rates for spheroidal figures of equilibrium. \cite{jacobi1834ueber} showed that a class of triaxial ellipsoids are also solutions of the problem (in this case, the fluid rotates as a rigid body, which makes possible to omit viscosity). It was shown later \cite[][]{meyer1842aequilibrii,liouville1846sur,liouville1855formules} that the Maclaurin spheroids become unstable above a certain deformation (fig. \ref{fig:rieman}), bifurcating into triaxial ellipsoids \cite[see][for historical details]{lyttleton1953stability}. More specifically, denoting by $a$ and $c$ the equatorial and polar radii, a dynamic instability appears at $c/a \approx 0.30$, whereas the secular instability appears at $c/a \approx 0.58$, where the triaxial Jacobi ellipsoid solutions branch on (in the limit of circular equator). Adding an internal uniform vorticity leads to consider two parts in the fluid velocity: an angular velocity of rigid-body rotation, and a motion of uniform vorticity superimposed on the latter. Each of these motions can be characterized by a three-component vector (time-dependent, in the general case). 
% Then, it can be shown that interchanging these vectors provides a different solution of the equations for which the geometric figure is the same (i.e. the semi-axes of the ellipsoid are identical): this is Dedekind's theorem \cite[][]{dirichlet1860untersuchungen,dedekind1861zusatz}. These ellipsoids are said to be {adjoined} to each other \cite[see e.g.][for details]{chandrasekhar1969ellipsoidal}. An example of such a pair of adjoint configurations is the Jacobi-Dedekind pair: the Jacobi ellipsoid is at rest in frame of reference rotating with a constant angular speed and the Dedekind ellipsoid is at rest in the inertial frame but with a fluid velocity of uniform vorticity. 
\cite{riemann1860untersuchungen} solved the general problem of the figures of equilibrium in this case (ellipsoids in the zone between the uppermost and the lowest solid black lines in figure \ref{fig:rieman}), and discussed the stability of the steady state solutions, which are now usually referred to as the Riemann ellipsoids \cite[see also][for details]{chandrasekhar1969ellipsoidal}.

\begin{figure}
  \begin{center}
     \includegraphics[scale=0.45]{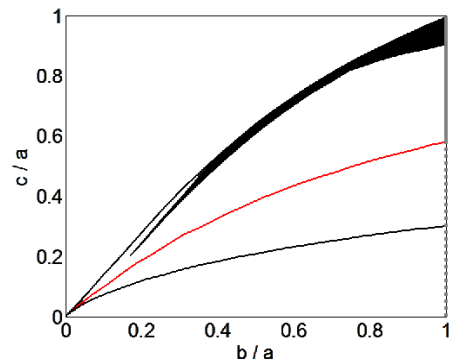}
    \caption{(Colour online) The local analysis of \cite{lebovitz1996new} shows that the only stable Riemann (inviscid) ellipsoids are actually in the small black zone in the ellipsoid aspect ratio planes. Red and black curves are the same as in figure 1.}
    \label{fig:rieman2}
  \end{center}
\end{figure}

Naturally, the stability of triaxial ellipsoids with internal uniform vorticity is directly related with the elliptical instability. However, the stability analysis of Riemann considers only perturbations that are linear polynomials of space coordinates and the energy criterion used by Riemann has been shown to be erroneous by \cite{lebovitz1966riemann}: \cite{riemann1860untersuchungen} {found} that the unstable ellipsoids for this perturbation are located between the solid blue line and the black solid upper line in figure \ref{fig:rieman}, whereas the correct unstable zone is actually smaller, given by the dark gray (blue) area in figure \ref{fig:rieman}, as shown by \cite{chandrasekhar1965equilibrum,chandrasekhar1966equilibrium} who considered also other kinds of perturbations (e.g. quadratic perturbations, which lead to unstable ellipsoids in the light gray (green) zone of figure \ref{fig:rieman}). The link with the elliptical instability has been made by \cite{lebovitz1996new,lebovitz1996short}, using a local analysis. These previously undetected elliptical instabilities affect most of the parameter space, and have rather large growth rates for the Dedekind family and nearby figures (fig. \ref{fig:rieman2}). The presence of the elliptical instability in these purely rotating isolated flows (absence of any tidal forces) is finally confirmed by \cite{ou2004nonlinear,ou2007further} with 3D numerical simulations of an inviscid compressible fluid. 
% \cite[even if the existence of compressible counter-parts of Riemann ellipsoids has been questioned by][]{chambat1994non,filippi1996implications,filippi2002functional}. 
These simple self-gravitating stellar models are thus subject to the elliptical instability which can then grow on fully compressible fluid flows.

The presence of the elliptical instability has been extensively studied in telluric bodies of the solar system as well as in Super-Earths \cite[][]{Cebron_2012}. However, its presence in gaseous bodies raises important issues. In this work, we address several points, as first steps toward a more complete study.

The paper is organized as follows. Section \ref{sec1} is devoted to the description of the problem. We also introduce the numerical procedure used for the simulation of the fluid flows in a triaxial ellipsoidal geometry. In section \ref{sec:sync}, we show how the elliptical instability is slightly affected by a latitudinal differential rotation with or without the presence of a synchronized latitude. In section \ref{sec:amp} we simulate the elliptical flow in a configuration close to a gaseous body and investigate the influence of the density profile and compressibility on a non-stationary mode. The threshold and the mode amplitude are characterized for a polytropic density profile and compared to the results obtained for a homogeneous fluid. Then, section \ref{sec:obs} raises the question of the observational signatures of the presence of the elliptical instability in hot-Jupiter systems as the spin-orbit misalignment. Having found that the instability may be present in stellar interiors, we consider in section \ref{sec:jup_chaud} the consequences of the angular momentum conservation. More precisely, we theoretically tackle the influence on the hot-Jupiter orbit of a change in the orientation of the stellar rotation axis when the elliptical instability is excited (leading to a spin-orbit misalignment). This is ideed important to check that the flows driven by the instability do not have not observed consequences on the orbit of the hot-Jupiter.

\section{Numerical model and theoretical basis} \label{sec1}

\subsection{Model} \label{sec:model}

\begin{figure}
  \begin{center}
     \includegraphics[scale=0.65]{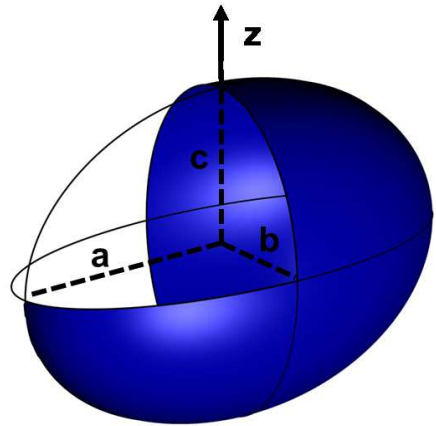}
    \caption{Sketch of the problem under consideration : a flow is rotating along the z-axis inside a triaxial ellipsoid of axes $(a,b,c)$.}
    \label{fig:scheme}
  \end{center}
\end{figure}

Let us consider the star as a massive fluid body, with its own spin rate and an imposed density profile. The star is tidally deformed by an orbiting extra-solar planet. In this study, the considered layer of the star is modeled by a rotating flow enclosed in a rigid triaxial ellipsoid of axes $(a,b,c)$, where $c$ is the
length of the polar axis and $a$ the major equatorial axis (figure \ref{fig:scheme}). The whole ellipsoid is then rotating at the rotation rate of the tides, i.e. at the orbital rotation rate $\Omega_{orb}$. We define the length-scale as the mean
equatorial radius $R_{eq}=(a+b)/2$ and the timescale, $\Omega^{-1}$, is defined using the tangential velocity along
the deformed outer boundary at the equator $U=\Omega \,R_{eq}$. The hydrodynamic problem is fully described by 4 dimensionless numbers: the ellipticity $\beta=|a^2-b^2|/(a^2+b^2)$, the aspect ratio $c/a$, the dimensionless orbital rotation rate $\Omega_{orb} / \Omega$ and the Ekman number $E=\mu/(\rho_{max} \Omega\,{R_{eq}}^2)$, where $\mu$ is the dynamic viscosity of the fluid (and $\rho_{max}$ the maximum density in the case of variable density).

\subsection{Growth rate of the instability}

The elliptic instability comes from a parametric resonance and can be excited in rotating fluids
whose streamlines are, even slightly, elliptically deformed. In the usual hydrodynamic situation, the underlying strain
field responsible for the elliptic deformation is stationary and in resonance with two inertial waves \cite[][]{bayly1986three,waleffe1990three}.  The same resonance mechanism may apply to magneto-inertial waves in the presence of an external magnetic field \cite[][]{lebovitz2004magnetoelliptic,mizerski2009magnetoelliptic,mizerski2011influence}, and to gravito-inertial waves when a density stratification is present \cite[][]{le2006thermo,guimbard2010elliptic}. However, to simplify the problem, we neglect the effects of a possible density stratification and the presence of a magnetic field \cite[see also][for a discussion on these specific points]{Cebron_2012}. Note also that if the flow is unstable in the absence of convection, then it is unstable in the presence of convection for the typical ranges of thermal Rossby number reached in stars and gaseous planets \cite[][]{lavorel2010experimental,cebron2010tidal}.

In all natural configurations such as binary stars, moon-planet or planet-star systems, orbital
motions are also present. Thus, the gravitational interaction responsible for the tidal deformation is rotating with
an independent angular velocity, $\Omega_{orb}$, different from the spin. This significantly {changes} the
conditions for resonance and the mode selection process, as recently studied in \cite[][]{le2007coriolis,le2010tidal}. In this situation, an important point is the appearance of a so-called 'forbidden zone' for $-1 \leq \Omega / \Omega_{orb} \leq 1/3$. For these parameters, resonances are not allowed \cite[e.g.][]{le2010tidal,Cebron_2012}. Outside this forbidden zone, the growth rate of the elliptical instability is given by \cite[see e.g.][]{Cebron_2012}.
\begin{equation}
\sigma=\frac{(2\ \Omega^G+3)^2}{16\ |1+\Omega^G|^3}\ \beta -K\ E^{\gamma}, \label{eq:sigsimpl_OHP}
\end{equation}
%%%% SImplifier l expression (partie num\"{\i}?`�rique, etc...) - en fait simplifier la partie en dessous, c est un peu brouillon %%%% AS
with $\Omega^G=\Omega_{orb}/(\Omega-\Omega_{orb})$, where $\Omega_{orb}$  is the orbital rate and $\Omega$ the fluid angular velocity. In this equation, the first term on the right-hand side comes from the inviscid
local analysis. The second term is due to the viscous damping of the instability. The
pre-factor $K$ depends on the excited mode of the elliptical instability, whereas $\gamma$ depends on the viscous damping. When surface damping is dominant, as in the case of
no-slip boundaries, $\gamma$ is equal to $1/2$. On the contrary, without
surface viscous damping, the bulk damping in the bulk is dominant and $\gamma=1$. In planetary cores
or laboratory experiments, no-slip boundary conditions apply, and the surface damping is dominant. However, in
stars, the boundary conditions between the radiative and the convective zones are less clearly defined. The astrophysical literature shows a disagreement for the viscous pumping/dissipation. Indeed, \cite{tassoul1987synchronization,tassoul1995orbital}, {argue} that it scales as $E^{1/2}$ \cite[see also][]{tassoul1990time,tassoul1992comparative,tassoul1992efficiency,tassoul1997synchronization} and \cite{rieutord1992ekman,espinosa2007dynamics,rieutord2008dynamics}, {argues} that this Ekman pumping and dissipation is rather proportional to $E$ \cite[see also][]{rieutord1997ekman}. For an incompressible homogeneous fluid star, we consider in this paper that the viscous dissipation scales as $E$. However, we can also consider a more complex model with an Ekman layer at the boundary between the radiative and the convective zone due to a viscosity/density jump and velocity gradients (in this case, the viscous dissipation scales as $E^{1/2}$).

\subsection{Numerical method}

We implement the model described in section \ref{sec:model} in the numerical code COMSOL Multiphysics\textsuperscript{\circledR}. The effect of the Coriolis force, which basically selects given modes and {modifies} the threshold of the instability, has already been numerically characterized in \cite{cebron2010systematic}. Therefore, in the numerical simulations presented in this work, we consider an ellipsoidal domain at rest in the inertial frame and we solve the following dimensionless compressible fluid dynamics equations:
\begin{eqnarray}
\rho \left[ \frac{\partial \mathbf{u}}{\partial t}+  \mathbf{u} \cdot
\mathbf{\nabla} \mathbf{u} \right] &=& -\mathbf{\nabla} p + E\mathbf{\nabla}^2 \mathbf{u}, \\
\frac{\partial \rho}{\partial t}+\mathbf{\nabla}  \cdot( \rho \,\mathbf{u}) &=&0, \label{eq:divnon}
\end{eqnarray}
where $\mathbf{u}$ is the velocity in the inertial reference frame, $p$ the pressure, and $\rho$ is the fluid density normalized by its maximum value. The applied boundary conditions are either an imposed tangential velocity in section \ref{sec:sync}, or a so-called compromise \textit{ad hoc} stress-free condition in section \ref{sec:amp}. 
Note that the fluid is assumed to be a Stokes fluid, i.e. the volume viscosity - or bulk viscosity -  is zero \cite[this is an usual hypothesis, see e.g.][]{jones2011anelastic}. {Thus}, the mechanical power in the bulk associated to viscous forces is only due to shears and independent of volume variations. Note also that the density profile is imposed at each {time step} in the calculations. This hypothesis simply means that the simulated flows do not modify the mean density profile of the modeled astrophysical body. Nevertheless, an energy cost for radial motions is taken into account  through equation (\ref{eq:divnon}). 

The mesh element type used for the fluid variables is the standard Lagrange element
$P1-P2$, which is linear for the pressure field and quadratic for
the velocity field. For time-stepping, we use the Implicit Differential-Algebraic solver (IDA solver), based on variable-coefficient Backward Differencing Formulas or BDF \cite[see][for details on the IDA solver]{hindmarsh2005sundials}. The integration method in IDA is variable-order, the order ranging between 1 and 5. At each time step the system is solved with the sparse direct linear solver PARDISO (www.pardiso-project.org). The validation of this numerical procedure to study the elliptical instability has been performed in \cite{cebron2010systematic}.

\section{Influence of a stellar differential rotation on the elliptical instability} \label{sec:sync}

$\tau$ Boo is a star characterized by a strong differential rotation \cite[][]{fares2009magnetic}, i.e. the angular velocity of the flow significantly varies from the equator ($\sim 2.1\ \textrm{rad.d}^{-1}$) to the poles ($\sim 1.6\ \textrm{rad.d}^{-1}$). Moreover, at the latitude $\theta_{sync}=38^{\circ}$, the surface layer of $\tau$ Boo rotates in $3.31$ days, which is equal to the orbital period of the hot Jupiter $\tau$ Bootis. It has been suggested that the peculiar magnetic field dynamics of $\tau$ Boo  is related to tides via a possible excited elliptical instability \cite[][]{cebron2010ohp}. However, it remains {unknown} if the elliptical instability can grow on this kind of base flow. In this section we characterize the presence of an elliptical instability on such a base flow for an incompressible fluid ($\rho=\textrm{cte}$).
\begin{figure}
  \begin{center}
     \includegraphics[scale=0.45]{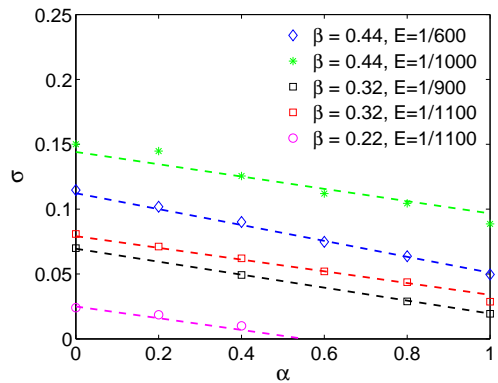}
    \caption{(Colour online) Growth rate of the elliptical instability vs. the amplitude of the differential rotation, in the absence of a synchronized latitude ($c=(a+b)/2$). Symbols stand for the results of the numerical simulations, and show good agreement the theoretical expression (\ref{eq:diff1}) (dashed lines).}
    \label{fig:diff}
  \end{center}
\end{figure}
In the inertial frame of reference, the model of differential rotation of \cite{fares2009magnetic} leads to the following dimensionless rotation rate:
\begin{eqnarray}
\omega=1- \alpha \sin^2 \theta, \label{eq:diff}
\end{eqnarray}
where $\omega$ is the dimensionless angular velocity at the latitude $\theta$, and $\alpha$ represents the difference in the rotation rate between the pole and the equator (for instance, $\alpha \approx 1/4$ for $\tau$ Boo and $\alpha \approx 1/5$ for the Sun). In the frame of reference rotating with the orbit of the hot-Jupiter, this angular velocity is written $1-\Omega_{orb}/\Omega-\alpha \sin^2 \theta$. We need to consider two different cases: the presence of a synchronized latitude, $\theta_{sync}$, where $1-\Omega_{orb}/\Omega=\alpha \sin^2 \theta_{sync}$ and when there is no synchronized latitude, $1-\Omega_{orb}/\Omega >\alpha$.
%%%% A verifier pour le gamma %%% AS

\subsection{Differential rotation without synchronized latitude}

First, let us consider the absence of a synchronized latitude. The angular
velocity profile (\ref{eq:diff}) is imposed at the outer boundary.
Then, the usual base flow along elliptical streamlines is recovered
in each plane orthogonal to the rotation axis, with a rotation rate
changing with the latitude. Therefore, the inviscid growth rate of
the elliptical instability should remain valid, but one can expect a
supplementary viscous damping related to latitudinal variations.

This is confirmed by the numerical results shown in figure
\ref{fig:diff}. There, the excited mode is the spinover mode, which
is stationary at saturation. Its growth rate is linearly damped when
$\alpha$ is increased. Rather than the general expression of the
growth rate (\ref{eq:sigsimpl_OHP}), we can use the more accurate
formula specific to the growth rate of the spinover mode
\cite[e.g.][]{cebron2010systematic}:
\begin{eqnarray}
\sigma=\sqrt{-\frac{(a^2-c^2)(b^2-c^2)}{(a^2+c^2)(b^2+c^2)}}-(K_1+K_2 \alpha) \,\sqrt{E} \label{eq:diff1}
\end{eqnarray}
where $K_1$ is the surface viscous damping coefficient
\cite[see
e.g.][]{kudlick1966transient,lacaze2005elliptical,Cebron_2012} in
absence of differential rotation (for instance $K_1=2.62$ for the
spinover mode in the limit of small ellipticity), and $K_2=1.5$ is the numerically obtained
coefficient for the damping due to the differential rotation. Note
that the imposed velocity at the boundary leads to the presence of
an Ekman layer, which is likely not present in stars. Here, we want
to validate that the only influence of the differential rotation is
a slight change in the viscous damping: it does not totally prevent
the growth of the instability for the imposed boundary flow case
considered here. Note however that differential rotation in the
fluid interior may have a different effect.

The spinover mode numerically studied above is the most unstable mode in a fluid sphere. All along this paper, this mode will be considered as a paradigm for all possible other modes of the elliptical instability. However, even in the limit $E \ll 1$, kinematic and geometric constraints may select another mode of the elliptical instability, which would be probably non-stationary. In the absence of specific numerical simulations, extension of our results to a more general case remains speculative. But for all modes, the physics of the elliptical instability remain identical. Hence we expect that in all cases, the influence of the differential rotation is limited to an additional dissipative effect which, in the limit of small Ekman numbers relevant for stars, will not affect the growth of the elliptical instability.

\begin{figure}
  \begin{center}
    \begin{tabular}{ccc}
      \subfigure[]{\includegraphics[scale=0.29]{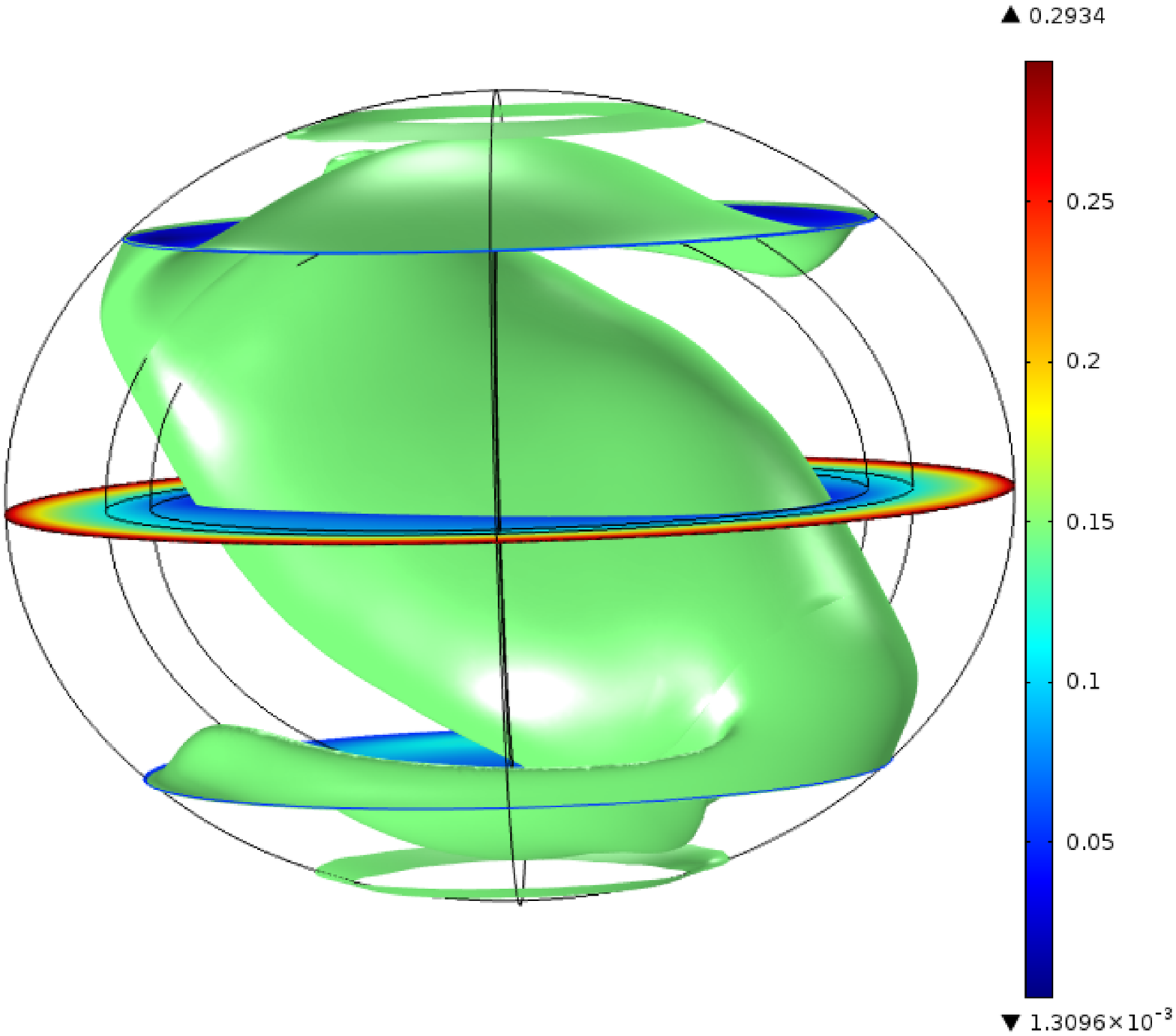}}\\
      \subfigure[]{\includegraphics[scale=0.4]{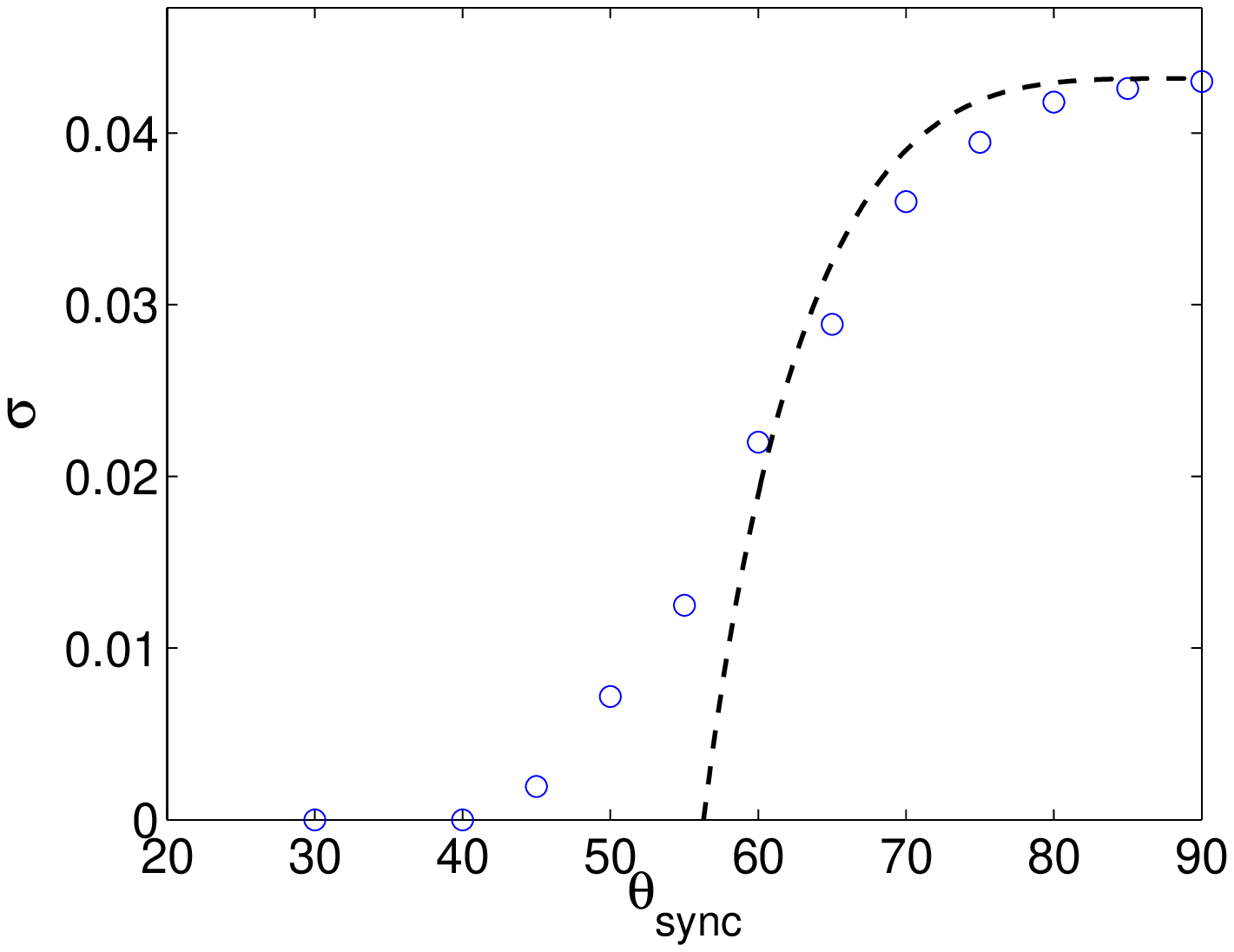}}
     \end{tabular}
\caption{(Colour online) Influence of the presence of a synchronized latitude ($c=\sqrt{(a^2+b^2)/2}$, $\alpha=1/2$, $E=1/2500$). (a) Visualization of the spinover mode for $\theta_{sync}=\pm 50^{\circ}$, with the velocity magnitude on the equatorial slice and slices corresponding to $\theta_{sync}$, and an isovalue surface $||\mathbf{u}||= 5 \%$ (surface created by all the points having a velocity magnitude of $5 \%$ of the equatorial boundary velocity). The unstable flow is confined between the synchronized latitudes. (b) Growth rate of the elliptical instability vs. the synchronized latitude. The numerical results (circles) show a relatively good agreement with relation (\ref{eq:sigma_c2}) and $c=\sin\theta_{sync}$, given by the dashed line.}
    \label{fig:diff2}
  \end{center}
\end{figure}

\subsection{Differential rotation with a synchronized latitude}

We now consider the presence of a synchronized latitude $\theta_{sync}$. In this case, we have
\begin{eqnarray}
\omega=\alpha\, (\sin^2 \theta_{sync}- \sin^2 \theta) \label{eq:synch_lat}
\end{eqnarray}
The numerical results presented in figure \ref{fig:diff2} show that, for the considered range of parameters, the elliptical instability can be excited when the synchronized latitude is large enough. According to the angular velocity profile (\ref{eq:synch_lat}), the relevant timescale becomes $\alpha\, \Omega$. Therefore, equation (\ref{eq:diff1}) for the growth rate of the spinover mode can be written:
\begin{eqnarray}
\sigma=\alpha \,\sqrt{-\frac{(a^2-c^2)\,(b^2-c^2)}{(a^2+c^2)\,(b^2+c^2)}}-2.62 \,\sqrt{\frac{E}{\alpha}}. \label{eq:sigma_c2}
\end{eqnarray}
Figure \ref{fig:diff2}(a) shows the results of a numerical simulation with a synchronized latitude $\theta_{sync}=\pm 50^{\circ}$ for an incompressible fluid ($\rho=\textrm{const}$). As seen in this visualization, the spinover mode is confined by the synchronized latitude, where the imposed velocity at the boundary is zero. This confinement effect is similar to the oblateness effect and we can model it using expression (\ref{eq:sigma_c2}) where $c$ is replaced by the height $c=\sin \theta_{sync}$ of the synchronized latitude. This leads us to consider an elliptical instability between the synchronized latitude and the equatorial plane, which is the most unstable plane due to the maximum angular velocity reached here. In this interpretation, the zone where the elliptical instability can be excited is thus smaller when the synchronized latitude $\theta_{sync}$ decreases. As a consequence, the growth rate of the elliptical instability gets larger (resp. smaller) when $\theta_{sync}$ increases (resp. decreases). In figure \ref{fig:diff2}(b), this estimation is compared with numerical results for the growth rate. A fairly good agreement is obtained, which confirms this confinement effect.

In astrophysics, we similarly expect that the elliptical instability will disappear around the synchronised latitude. Nevertheless, in the limit of small Ekman numbers reached in stars, we expect that the elliptical instability with a wavelength very small compared to $2 R_{eq} \theta_{sync}$ may be excited. Equation (\ref{eq:sigma_c2}) being specific to the spinover mode, for another mode, the growth rate is then simply given by equation (\ref{eq:sigsimpl_OHP}) rewritten in terms of the relevant timescale $\alpha\, \Omega$.

\section{Influence of a density profile on a non-stationary mode} \label{sec:amp}

In this section, we present the first numerical simulations of the elliptical instability in ellipsoids with the compromise \textit{ad hoc} boundary conditions originally suggested by \cite{mason1999nonlinear}. These boundary conditions (BC), that will be presented below, provide an efficient way to study the influence of the streamlines deformation without viscous Ekman layers at boundaries. The amplitude of the elliptical instability has been numerically shown to be of order $1$ in \cite{cebron2010systematic} for the spinover mode but the generality of this result remains to be verified. Here we focus on the non-stationary mode (1,3), excited in oblate ellipsoids, where the polar axis is the smallest one. This mode (1,3) corresponds to the resonance of two modes with azimuthal numbers $m=1$ and $m=3$ with the base flow \cite[e.g.][for details]{cebron2010systematic}. We first confirm that the scaling laws obtained for the spinover mode remain valid for mode (1,3). Then, we impose a polytropic density profile to study the influence of a mass concentration on the saturation of the elliptical instability.

\subsection{Simulation of the elliptical instability in a gaseous bodies}

The ideal BC for a gaseous body would allow the surface of the fluid domain to deform under the action of the flow, maintaining the rotation rate of the fluid. It is a challenging numerical task. {Indeed, solving the deformations of a fluid ellipsoidal domain requires us to use a deformable mesh (e.g. with an Arbitrary Lagrangian Eulerian method), or to use methods developed for diphasic flows such as front capturing  (implicit representation of the interface as in the Volume-Of-Fluid method or the level-set method) or front tracking methods (explicit representation of the interface), or to use meshless methods (e.g. smoothed-particle hydrodynamics method)}. These complex and computationally expensive methods should then be used with stress-free BC at the surface of the star, which raises problems to maintain the rotation rate \cite[e.g.][]{wu2009dynamo}. Here, we choose as a first step to use the simple BC proposed by \cite{mason1999nonlinear} on a fixed fluid domain: only the difference $\mathbf{\tilde{u}}$ with the base flow is solved, and stress-free BC is imposed on $\mathbf{\tilde{u}}$. Writing the velocity $\mathbf{u}= \mathbf{u_b} + \mathbf{\tilde{u}}$, the velocity $\mathbf{\tilde{u}}$ represents the secondary motions that are excited when the base flow $\mathbf{u_b}=-ay/b\ \mathbf{e_x}+bx/a\ \mathbf{e_y}$ is unstable (where the unit vectors $\mathbf{e_x}$ and $\mathbf{e_y}$ are given by the two axes of the elliptical equator). The problem is then described by the system of dimensionless equations:
\begin{eqnarray}
\frac{\partial \mathbf{\tilde{u}}}{\partial t}+ \mathbf{\tilde{u}} \cdot
\mathbf{\nabla} \mathbf{\tilde{u}} &=& -\mathbf{\nabla} p + E\
\mathbf{\nabla}^2 \mathbf{\tilde{u}} +\mathbf{f}, \\
\frac{\partial \rho}{\partial t}+\mathbf{\nabla}  \cdot( \rho\ \mathbf{\tilde{u}}) &=&0. \label{eq:diver}
\end{eqnarray}
where $\mathbf{f}=- \mathbf{u_b} \cdot \mathbf{\nabla} \mathbf{\tilde{u}}- \mathbf{\tilde{u}} \cdot \mathbf{\nabla} \mathbf{u_b}$, and with stress-free BC for $\mathbf{\tilde{u}}$ (eq. \ref{eq:diver} is verified by the basic flow $\mathbf{u_b}$).
\begin{figure}
  \begin{center}
     \includegraphics[scale=0.45]{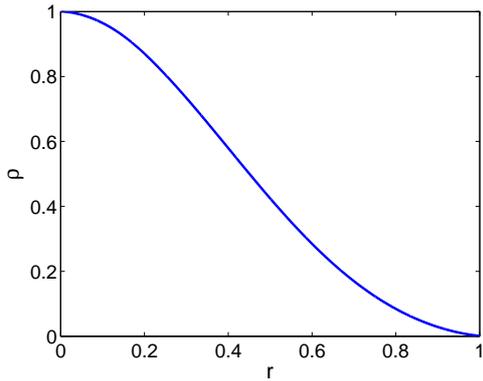}
    \caption{Density profile imposed at each time step along ellipsoids homothetic to the surface $r=\sqrt{(x/a)^2+(y/b)^2+(z/c)^2}=1$.}
    \label{fig:poly_density}
  \end{center}
\end{figure}
To investigate the influence of compressibility, we impose at each time step the density profile given in figure \ref{fig:poly_density}, which corresponds to a polytropic density profile of index $n=3/2$, varying between $1$ at the center to $10^{-3}$ at the outer surface, such that the density is constant along ellipsoids which are homothetic to the surface, i.e. ellipsoids with the same center and the same axes ratio \cite[see][]{lai1993ellipsoidal}.
We use the bulk force $\rho \mathbf{g}$, where the gravity $\mathbf{g}=-  \mathbf{\nabla} \phi$ is given by the dimensionless Poisson equation $\bigtriangleup \phi = 3\ \rho$ for the gravitational potential $\phi$ ($\bigtriangleup$ being the Laplacian operator), assuming that the gravity is equal to unity at the surface of a sphere of radius $1$. We impose, as a boundary condition, that the outer boundary is an isopotential \cite[see][for details]{cebron2010tidal}.

\subsection{Threshold}
% \begin{figure}
%   \begin{center}
%      \includegraphics[scale=0.45]{freq_poly.eps}
%     \caption{Frequency of the flow driven by mode (1,3) of the elliptical instability at saturation as a function of the threshold distance for an homogeneous fluid ellipsoid (circles, diamonds) and a polytrope 3/2 fluid (*). The scaling law in $\omega=\delta^{-1/2}$, represented by the solid black line, is in good agreement with the numerical results for an homogeneous fluid ellipsoid.}
%     \label{fig:sig_freq}
%   \end{center}
% \end{figure}
In this section, we focus on the amplitude of an non-stationary mode of the elliptical instability, the mode (1,3) \cite[see][]{cebron2010systematic}. This mode is excited when the polar radius is the smallest axis and $\Omega_{orb}=0$. We first consider the incompressible flow driven by mode (1,3). The BC used here do not generate Ekman layers, and the viscous damping is thus essentially a bulk damping, which scales as $E$. A WKB (Wentzel-Kramers-Brillouin) stability analysis shows that the growth rate can be estimated in the limit $\beta \ll 1, E \ll 1$ by \cite[e.g.][]{ledizes2000three}
\begin{eqnarray}
\sigma=\frac{9}{16} \beta-k^2 E, \label{eq:sig_free}
\end{eqnarray}
where $k$ is the dimensionless wave-number of the excited mode. It leads to thresholds very different from the case of no-slip boundary conditions where the presence of the viscous Ekman layers lead to a damping proportional to $\sqrt{E}$. For instance, for $\beta=0.317, c/a=0.65$, the critical Ekman number is around $E_c \approx 1/450$ for no-slip BC, whereas it is $E_c \approx 1/85$ with the BC used here. Expression (\ref{eq:sig_free}) shows that the relevant control parameter of the instability, for these boundary conditions, is $\chi=\beta/E$. Thus, we define the threshold distance by  $\delta=(\chi-\chi_c)/\chi_c$, where $\chi_c \sim k^2$ is the critical control parameter (i.e. the value of $\chi$ for $\sigma=0$). In the case of mode (1,3), $k=2 \pi$ \cite[][]{Cebron_mag2012}, and we thus expect a threshold around $\chi \approx 40$. Given the finite values of $\beta$ and $E$, this asymptotic estimation is in rather good agreement with the numerically obtained threshold of $\chi_c \approx 25$ for the homogeneous incompressible case. In contrast, the threshold is significantly increased in the polytropic case, where the critical control parameter is $\chi_c \approx 275$, i.e. $11$ times larger.
\begin{figure}
  \begin{center}
    \begin{tabular}{ccc}
      \subfigure[]{\includegraphics[scale=0.4]{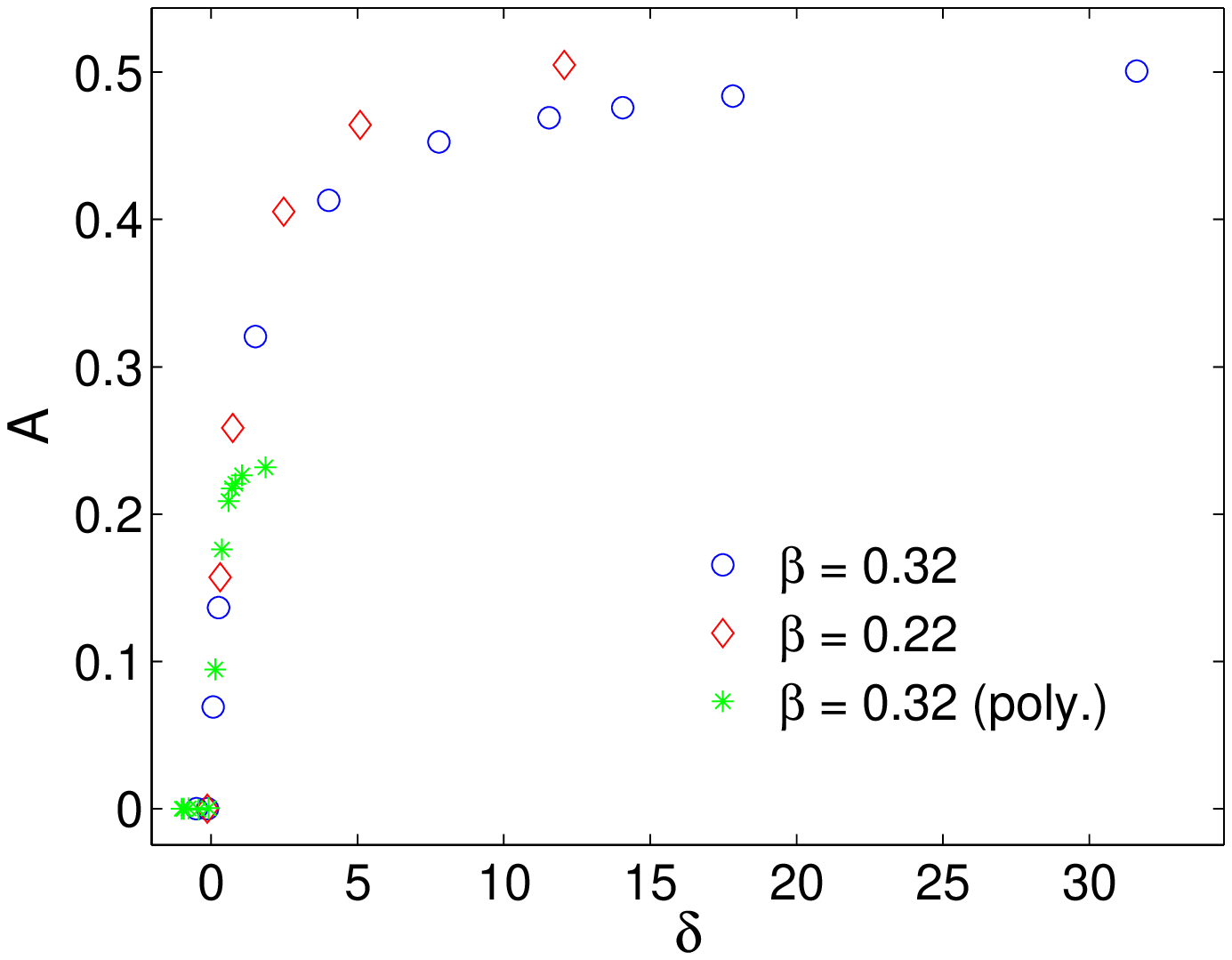}} \\
      \subfigure[]{\includegraphics[scale=0.4]{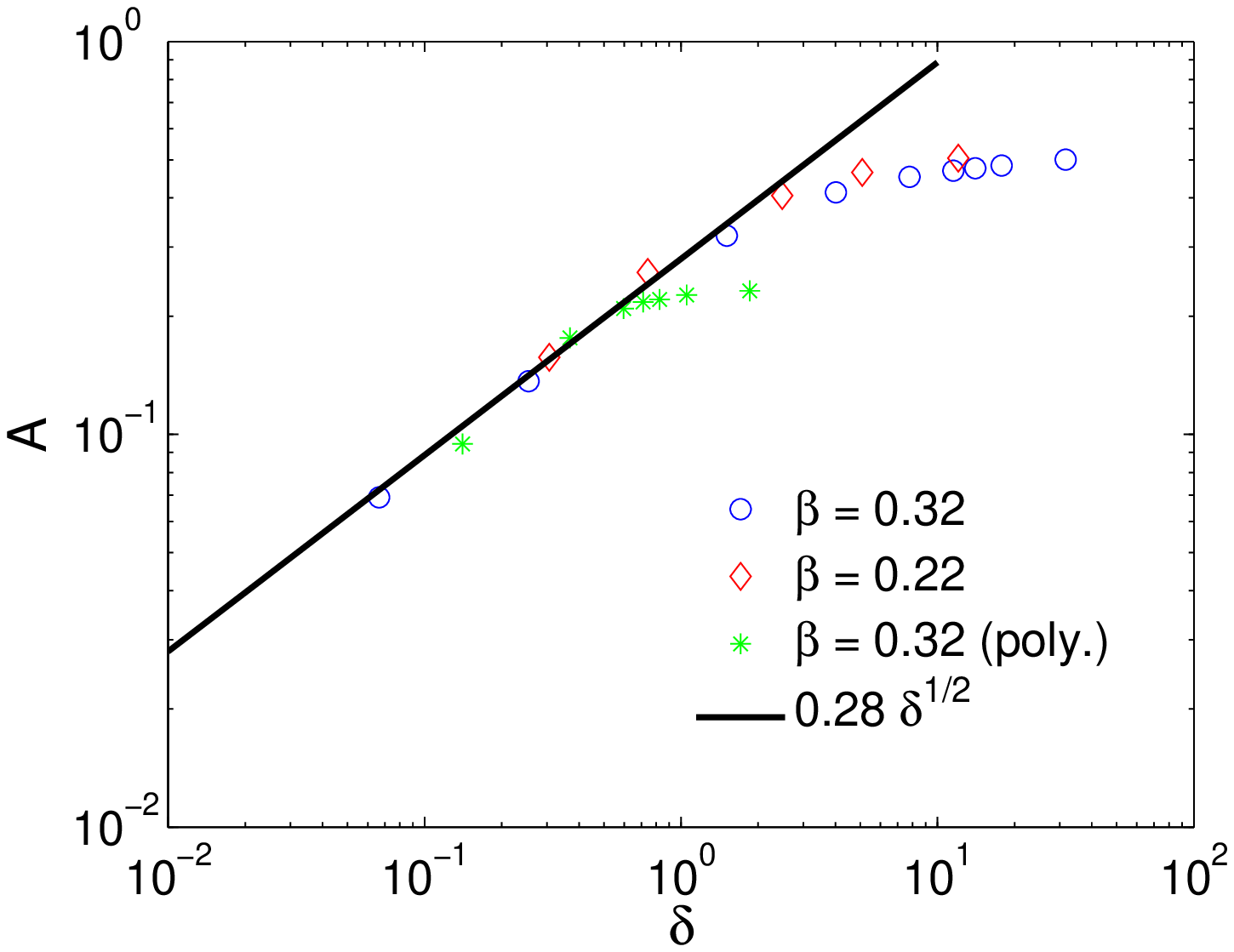}}
     \end{tabular}
\caption{(Colour online) (a) Amplitude at saturation of the flow driven by the mode (1,3) of the elliptical instability vs. the threshold distance for a homogeneous fluid (circles, diamonds) and a polytrope of index 3/2 (*). (b) Comparison with a scaling law in $\sqrt{\delta}$ close to the threshold.}
    \label{fig:amp}
  \end{center}
\end{figure}
% Mode (1,3) is an non-stationary mode which oscillates at a rate $\omega$ of about two times the rotation rate \cite[e.g.][]{cebron2010systematic}. Here, using a Fourier analysis of our temporal signals, we study the variation of this frequency with the threshold distance and in presence of a non-uniform density profile. In figure \ref{fig:sig_freq}, this pulsation of $\omega \approx 2$ is confirmed close to the threshold, but a clear viscous detuning of the frequency is obtained, following a scaling law in $\omega \sim \delta^{-1/2}$.

\subsection{Mode amplitude}

Since we directly solve the difference with the base flow, we can use a more accurate definition of the amplitude of the flow driven by the instability than the one used before \cite[e.g.][]{cebron2010systematic,Cebron_mag2012}. We define the amplitude $A$ of the elliptical instability as the time average value, at the saturation of the instability, of the volume root mean square velocity
\begin{eqnarray}
u_{rms}=\sqrt{ \frac{1}{V} \int_V \mathbf{u}^2 \textrm{d}v},
\end{eqnarray}
where $V$ is the volume of the fluid domain. Numerical results in figure \ref{fig:amp} show that the general conclusions of \cite{cebron2010systematic} remain valid for this different mode and boundary conditions, i.e. the amplitude of the elliptical instability scales as $\sqrt{\delta}$ near the threshold (the bifurcation remains supercritical) and we expect the amplitude to go towards a value around $1$ far from the threshold. Moreover, for a same threshold distance $\delta$, numerical results in figure \ref{fig:amp} show that the driven flows are of the same order of magnitude for the homogeneous case and the compressible simulation (the amplitude for the polytropic fluid seems to be smaller by a factor $2$ far from the threshold).

\section{Observational signatures of the elliptical instability?} \label{sec:obs}

Previous sections show that the elliptical instability can be expected in gaseous planets and stars even if its existence is difficult to detect. Another way to tackle the problem is to consider observational signatures of the elliptical instability. For instance, the presence of an elliptical instability may have signatures on the stellar magnetic field \cite[][]{lacaze2006magnetic,lebarsNature}. In this work, we only consider as a first step preliminary results on two other possible signatures: the radii anomalies of hot Jupiter, which are tested here as a proxy of the presence of the elliptical instability in the planet, and the spin-orbit misalignment of certain stars as a proxy of the flow driven by the instability in the star.

\subsection{Tidal deformation}

For application of our conclusions to real systems, we need a good estimate of the expected tidal deformations.
We denote by $M$ and $R$ the mass and the radius of the star, respectively, $M_2$ the mass of the body responsible for the
tidal deformation and $D$ the distance between the two. The ellipticity due to tidal forces can be estimated \cite[e.g.][]{Cebron_2012}
\begin{eqnarray}
\beta=\frac{3}{2}\ h_2\ \frac{M_2}{M}\frac{R^3}{D^3} \label{eq:newton3}
\end{eqnarray}
with the radial displacement Love number $h_2$, directly linked to
the potential Love number $k_2$ by $h_2=1+k_2$ (note that this is only true for a body in hydrostatic equilibrium for which elastic stress is negligible; \cite{leconte2011distorted}). A lower bound of the tidal deformation is thus given by $k_2=0$, which is the limiting case of a massless tidal bulge. Taking into account the gravitational potential of the tidal bulge systematically increases the tidal deformation. This influence is completely encompassed into $k_2$, which depends only on the density profile at first order \cite[][]{leconte2011distorted}. This density profile can be obtained by assuming that the star is made of a polytropic gas fluid, which is a commonly used approximation. In this model, the polytropic index $n$ ranges typically between $n=3/2$, representative of a fully convective star, and $n=3$, which gives a stably stratified polytropic fluid and is thus representative of radiative zones \cite[e.g.][]{horedt2004polytropes}. For these two limiting polytropes, the potential Love numbers are respectively $k_2 \approx 0.3$ and $k_2 \approx 0.03$ \cite[][]{brooker1955apsidal}. According to \cite{mazeh2007observational}, typical model of stars leads to $k_2 \in [0.001; 0.01]$, depending on the stellar mass and age. In order to be conservative, we consider the lower bound $k_2=0$ \cite[see also][for estimations of $k_2$ from the apsidal motions constants]{claret2002new,landin2009combined}.

\begin{table*}
\caption{Physical and orbital characteristics of the extra-solar planets considered in this work.}
\label{cebronfig8}
%\begin{center}
{\scriptsize
%{\scriptsize
\begin{tabular}{lcccccccccccc}
\\
\hline
Stars  & $M_1$ & $R_1$ & $M_2$ & $R_2$ & $e$ & $T_{orb}$ & $\beta$ & $\Delta_R$ & $\langle \tau \rangle$ \\
$ $  & $(M_s)$ & $(R_s)$ & $(M_j)$ & $(R_j)$ & $ $ & $(d)$ & $(\times 10^3)$ & $ $ & $(years)$ \\
% $ $ & $ $ & $ $ & $ $ & $ $ & $ $ & $ $ & $ $ & $ $ & $ $ & $ $ \\

\hline
WASP-17b  & 1.20 & 1.38 & 0.49 & ${}^1$ 1.0 & 0.13 & -3.73 & 3.39 & $0.575 \pm 0.26$ & 20 \\
WASP-10  & 0.75 & 0.70 & 3.15 & 1.08 & 0.06 & 3.09 & 0.94 & $-0.049 \pm 0.047$ & 364 \\
CoRoT-4 & 1.16 & 1.17 & 0.72 & 1.19 & 0.00 & 9.20 & 0.63 & $0.045 \pm 0.063$ &  - \\
WASP-19  & 0.96 & 0.94 & 1.15 & 1.31 & 0.00 & 0.79 & 65 & - &  - \\
WASP-18  & 1.25 & 1.22 & 10.30 & 1.11 & 0.00 & 0.94 & 3.05 & $0.097 \pm 0.1$ &  - \\
CoRoT-6  & 1.05 & 1.02 & 2.96 & 1.17 & 0.00 & 8.89 & 0.16 & - &  - \\
kepler-5b  & 1.45 & 2.07 & 2.16 & 1.68 & 0.00 & 3.55 & 3.85 & - &  - \\
TrES-3  & 0.90 & 0.80 & 1.92 & 1.29 & 0.00 & 1.31 & 14.07 & $0.116 \pm 0.87$ &  - \\
HD41004b  & 0.40 & 0.40 & 19.00 & ${}^1$ 1.0 & 0.08 & 1.33 & 9.1 & - & 8.2 \\
Tau-boo  & 1.30 & 1.33 & 3.50 & ${}^1$ 1.0 & 0.00 & 3.31 & 0.65 & - &  - \\
HD179949  & 1.28 & 1.19 & 0.95 & ${}^1$ 1.0 & 0.02 & 3.09 & 2.52 & - & 1102 \\
CoRoT-7  & 0.93 & 0.87 & 0.01 & 0.15 & 0.00 & 0.85 & 7.0 & - &  - \\
GJ-674  & 0.35 & 0.35 & 0.04 & ${}^1$ 1.0 & 0.20 & 4.69 & 34.37 & - & 0.82 \\
HD17156b  & 1.24 & 1.44 & 3.21 & 1.02 & 0.67 & 21.21 & 0.16 & $-0.076 \pm 0.07$ & 12 \\
HD189733b  & 0.82 & 0.79 & 1.13 & 1.14 & 0.03 & 2.22 & 6.13 & $-0.011 \pm 0.04$ & 170 \\
HD209458b  & 1.01 & 1.15 & 0.69 & 1.32 & 0.07 & 3.52 & 5.68 & $0.20 \pm 0.03$ & 47 \\
HAT-P-1b  & 1.13 & 1.12 & 0.52 & 1.23 & 0.07 & 4.46 & 4.08 & $0.11 \pm 0.059$ & 92 \\
CoRoT-2b  & 0.97 & 0.90 & 3.31 & 1.47 & 0.00 & 1.74 & 7.07 & $0.26 \pm 0.09$ &  - \\
HD149026b  & 1.30 & 1.50 & 0.36 & 0.65 & 0.18 & 2.88 & 2.49 & - & 9.2 \\
TrES-2b  & 0.98 & 1.00 & 1.20 & 1.27 & 0.08 & 2.47 & 6.55 & $0.17  \pm 0.04$ & 21 \\
CoRoT-3b & 1.36 & 1.54 & 21.23 & 0.99 & 0.00 & 4.26 & 0.06 & $-0.023  \pm 0.07$ &  - \\
TrES-1b  & 0.87 & 0.82 & 0.61 & 1.08 & 0.00 & 3.03 & 5.04 & $-0.025  \pm 0.04$ &  - \\
HAT-P-2b  & 1.36 & 1.64 & 9.09 & 1.16 & 0.52 & 5.63 & 0.45 & $0.1  \pm 0.09$ & 3.6 \\
CoRoT-1b  & 0.95 & 1.11 & 1.03 & 1.49 & 0.00 & 1.51 & 31.7 & $0.26  \pm 0.09$ &  - \\
HAT-P-7b  & 1.47 & 1.84 & 1.80 & 1.42 & 0.00 & -2.20 & 7.31 & $0.27  \pm 0.14$ &  - \\
WASP-14b  & 1.32 & 1.30 & 7.72 & 1.26 & 0.09 & 2.24 & 1.19 & $0.08  \pm 0.08$ & 82 \\
XO-3b  & 1.21 & 1.38 & 11.79 & 1.22 & 0.26 & 2.19 & 0.47 & $0.19  \pm 0.07$ & 13 \\
HD80606  & 0.90 & 0.97 & 3.94 & 1.03 & 0.93 & 111.44 & 0.44 & $-0.049  \pm 0.02$ & 1.4 \\
\hline
\\
\end{tabular}  }
%\end{center}
{\scriptsize

$^1$ Radius unknown: we choose a default value of $1\ R_j$.\\

 The masses $M_1$ of the stars are given in Sun mass unity $M_s$, the masses of the exoplanets $M_2$ are given in Jupiter mass unity $M_j$, the radius $R_1$ of the stars is given in Sun radius $R_s$, the orbital periods are in days ($d$), the radius anomaly $\Delta_R$ is dimensionless and the typical (inviscid) average growth time $\langle \tau \rangle$ is in years. A value of $0$ for $e$ indicates a value beyond the detection limit ($e \lesssim 0.01 $). Note that the extra-solar planets with an orbital eccentricity equal to zero cannot be unstable by this mechanism.}
  \end{table*}

\subsection{The bulk viscous damping}
A usual WKB analysis shows that the pre-factor $K$ of the viscous damping term in
equation (\ref{eq:sigsimpl_OHP}) is simply the square of the dimensionless wave-number $k=2 \pi/
l_1=\sqrt{K}$ of the excited mode \cite[e.g.][]{Cebron_2012}. However, the value of $k$, needed to
calculate explicitly the damping term, changes with the excited
mode. A lower bound estimation for $k$ is given by $k=\pi$,
which corresponds to the first unstable mode (a half wavelength
inside the gap of the shell). A relevant upper bound is difficult to
obtain. However, the balance between the typical inviscid growth
rate $\sigma \sim \beta$ and the volume damping term leads to a typical dimensionless
wavelength $l_2 \sim \sqrt{E/\beta}$. Considering extra-solar planets, table \ref{cebronfig8} gives $\beta \sim 10^{-4}-10^{-2}$ and $E \sim 10^{-15}-10^{-13}$  \cite[considering a kinematic viscosity of $\nu=10^{-6}\ m^2\ s^{-1}$][]{zhang1996penetrative}, which leads to a typical scale of $l_2 \sim 10^{-7}-10^{-5}$. Considering now stars, table \ref{cebronfig3} show $\beta \sim 10^{-8}-10^{-5}$ and $E \sim 10^{-17}-10^{-13}$, which leads to a typical scale of $l_2 \sim 10^{-6}-10^{-3}$. Given that the dimensionless viscous dissipation scale $l_{\nu}$ of the Sun is around $l_{\nu} \sim 10^{-12}$ at the surface and  $l_{\nu} \sim 10^{-10}$ at the bottom of the convection zone \cite[][]{rincon2010thesun}, it appears that the different scales are well separated in the flow, allowing the elliptical instability to grow on large scales. When $k \sim O(1)$, the calculated
dissipative term is so small that it can be neglected. We can
thus consider that this case of free-slip boundaries leads to a growth rate
of the same order as the inviscid growth rate $\sigma$. {The
associated typical timescale $\tau=\sigma^{-1}$ is thus chosen to
represent the strength of the instability.} This model is correct provided that $\tau \gg \Omega_{orb}^{-1}$, which make it possible to consider a quasi-static situation, and $\tau \ll \tau_2$ where $\tau_2$ is the typical time of stellar evolution. Typically, for the systems considered in this work, tides circularize the orbit and affect the obliquity on timescales (respectively $~ 10^9$ yr. and $~ 10^6$ yr.) that are shorter than the system age \cite[e.g.][]{levrard2006tidal,leconte2010tidal}. Considering the synchronization time $\tau_2$, it follows that the presence of the elliptical instability is expected in the range $1\ \textrm{yr.} \ll\, \langle\tau\rangle\, \ll 10^6\ \textrm{yr}$.
\begin{figure}
  \begin{center}
     \includegraphics[scale=0.5]{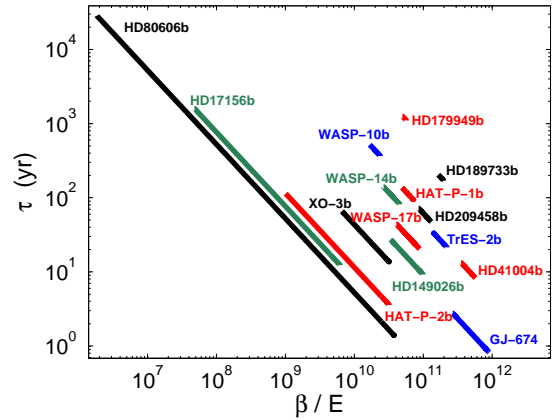}
    \caption{(Colour online) Typical (inviscid) growth time $\tau$ of the elliptical instability in the hot Jupiter vs. the control parameter $\beta/E$ (see eq. \ref{eq:sigsimpl_OHP}), assuming a molecular kinematic viscosity $\nu=10^{-6}\ m^2\ s^{-1}$ \cite[][]{zhang1996penetrative}.}
    \label{fig:obs0}
  \end{center}
\end{figure}

\subsection{Bloated hot Jupiter: elliptical instability causes extra-dissipation}

We now consider the gravitational tides raised by both the host star and the planet on each other. Following the traditional viscous approach of the equilibrium tide theory, \cite{hut1981tidal} shows that a typical hot Jupiter evolves to the spin rate value \cite[see also][]{leconte2010tidal}:
\begin{eqnarray}\label{131313}
\Omega=n\, \frac{1+15 \tilde{e}^2+30 \tilde{e}^4+5 \tilde{e}^6}{(1+6 \tilde{e}^2+2 \tilde{e}^4)(1-4 \tilde{e}^2)^{3/2}},
\end{eqnarray}
where $n$ is the orbital mean angular rate and $\tilde{e}=e/2$ is half the orbital eccentricity $e$. {Considering the age of the known transiting systems and the rather low angular momentum stored in the planet, the hot Jupiter is assumed to be of pseudo-synchronized, its spin rotation rate (\ref{131313}). For any value of $e$, equation (\ref{131313}) gives $\Omega > n$: hot Jupiters on eccentric orbit are expected to rotate faster than the orbital revolution, allowing the elliptical instability to develop in some of those cases (the ratio $T$ between the orbital period and the spin period is then larger than $1$ and thus no hot Jupiter is in the forbidden zone, defined by $T \in [-1; 1/3]$). Note, however, that extra-solar planets with an orbital eccentricity equal to zero cannot be unstable by the elliptical instability since there is no differential rotation between the flow and the tides.} Moreover, those synchronised hot Jupiters cannot undergo a libration driven elliptical instability such as Super-Earths \cite[][]{Cebron_2012}, which is related to physical libration of solid bodies.

\begin{figure}
  \begin{center}
     \includegraphics[scale=0.5]{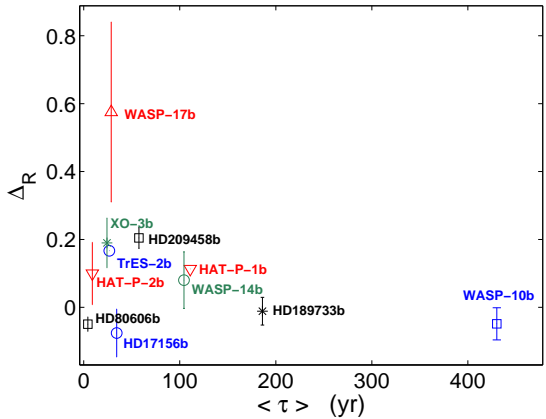}
    \caption{(Colour online) Radius anomaly $\Delta_R$ of hot Jupiters vs. the typical inviscid average growth time $\langle \tau \rangle$ of the elliptical instability in the planet. Error-bars represent the uncertainties on the radius anomaly.}
    \label{fig:obs}
  \end{center}
\end{figure}

All the hot-Jupiter system parameters needed for the calculations of this section are given in table \ref{cebronfig8}. Using the inviscid form of equation (\ref{eq:sigsimpl_OHP}), we  obtain an estimation of the typical growth time $\langle \tau \rangle$ of the elliptical instability in the hot Jupiter as shown in figure \ref{fig:obs0}. Because of the orbital eccentricity, this growth time is not constant along the orbit and one can thus define its mean value $\langle \tau \rangle$. In figure \ref{fig:obs}, $\langle \tau \rangle$ is represented as a  function of the radius anomaly $\Delta_R=(R_{obs}-R_{th})/R_{th}$, where $R_{obs}$ and $R_{th}$ are respectively the observed and the theoretical expected radii \cite[see][for the details of internal structure model used to estimate $\Delta_R$]{leconte2010tidal}. Following our hypothesis, the more unstable the elliptical instability is (i.e. the smaller $\langle \tau \rangle$ is), the stronger is the heat dissipation due to the vigorous associated driven turbulence, then the larger is the radius anomaly ($R_{th}$ does not take into account this supplementary heat dissipation). Results in figure \ref{fig:obs} are coherent with the presence of strong elliptical instabilities, which lead to high internal dissipation, and thus high $\Delta_R$. This trend has to be confirmed by considering a larger number of systems. Note that the very short growth time is due to the huge tidal deformation of the hot Jupiter and the absence of damping mechanism in our estimation. The results are quite robust. For instance, in the limit $E \ll 1$ reached in these systems, the presence of any Ekman layer would provide a dominant damping of the growth rate proportionnal to $\sqrt{E}$. Such a damping with a viscosity as high as $\nu=10^{-3}\, \textrm{m}^2.\textrm{s}^{-1}$ would change the stability of only one planet, WASP-10b (eq. (\ref{eq:sigsimpl_OHP}) with $\gamma=1/2$, $K=10$).

\begin{table*}
\caption{Physical and orbital characteristics used for the stability calculation. The upper (resp. lower) block corresponds to the $11$ (resp. $17$) considered stars outside of (resp. in) the forbidden zone.}
\label{cebronfig3}
%\begin{center}
{\tiny
%{\scriptsize
\begin{tabular}{lcccccccccccc}
\\
\hline
Stars &   $M_1$ & $R_1$ & $T_{spin}$ & $M_2$ & $T_{orb}$ & $T$ & $\beta$ & $E$ & $\eta$  & $\nu$ & $\lambda$ & $\langle \tau \rangle$ \\
$ $ &  $(M_s)$ & $(R_s)$ & $(d)$ & $(M_j)$ & $(d)$ & $ $ & $(\times 10^6)$ & $(\times 10^{16})$ & $ $  & $(cm^2\ s^{-1})$ & $(deg)$ & $(kyr.)$ \\
% $ $ & $ $ & $ $ & $ $ & $ $ & $ $ & $ $ & $ $ & $ $ & $ $ & $ $  & $ $ & $ $ \\

\hline
   CoRoT-4  & 1.16 & 1.17 & 9.40 & 0.72 & 9.20 & 0.98 & 1.95 & 1.66 & 0.84 & 8.5 &- & 3950 \\
CoRoT-6  & 1.05 & 1.02 & 6.40 & 2.96 & 8.89 & 1.39 & 0.71 & 1.56 & 0.81 & 9 & - & 43 \\
Tau-boo  & 1.30 & 1.33 & 3.50 & 3.50 & 3.31 & 0.95 & 9.3 & 0.46 & 0.88 & 8.1 & - & 12 \\
HD179949  & 1.28 & 1.19 & 7.60 & 0.95 & 3.09 & 0.41 & 2.1 & 1.26 & 0.87 & 8.3 & - & 63 \\
HAT-P-2  & 1.36 & 1.64 & 3.80 & 9.09 & 5.63 & 1.48 & 121 & 0.33 & 0.88 & 8.2 & - & 0.3 \\
HD17156b  & 1.24 & 1.44 & 48.0 & 3.21 & 21.2 & 0.44 & 7.24 & 5.84 & 0.81 & 8.9 & $5 \pm 5$ & 19 \\
HD209458  & 1.01 & 1.15 & 11.4 & 0.69 & 3.52 & 0.31 & 1.75 & 2.32 & 0.77 & 9.4 & $0.1 \pm 2.4$ & 451 \\
CoRoT-2  & 0.97 & 0.90 & 4.52 & 3.31 & 1.74 & 0.39 & 16.1 & 1.70 & 0.73 & 10.8 & $7.2 \pm 4.5$ & 7.4 \\
CoRoT-3  & 1.36 & 1.54 & 4.50 & 21.23 & 4.26 & 0.95 & 44 & 0.37 & 0.95 & 6.9 & $37.6 \pm 22$ & 3.3 \\
XO-3b  & 1.21 & 1.38 & 3.70 & 11.79 & 2.19 & 0.59 & 99 & 0.43 & 0.88 & 7.7 & $37.3 \pm 3.7$ & 0.4 \\
HD80606  & 0.90 & 0.97 & 44.0 & 3.94 & 111 & 2.53 & 18.4 & 16.1 & 0.87 & 12.1 & $42 \pm 8$ & 13 \\

\hline
WASP-17b  & 1.20 & 1.38 & 10 & 0.49 & 0.13 & -3.73 & -0.37 & 1.75 & 1.10 & 0.88 & - & 7.38 \\
WASP-10 & 0.75 & 0.70 & 11.90 & 3.15 & 0.06 & 3.09 & 0.26 & 4.5 & 11.82 & 0.65 & - & 17.14 \\
WASP-19  & 0.96 & 0.94 & 10.50 & 1.15 & 0 & 0.79 & 0.08 & 29 & 3.85 & 0.72 & - & 11.42 \\
WASP-18 & 1.25 & 1.22 & 5.60 & 10.30 & 0 & 0.94 & 0.17 & 231 & 0.84 & 0.88 & - & 7.82 \\
kepler-5  & 1.45 & 2.07 & 21 & 2.16 & 0 & 3.55 & 0.17 & 13.4 & 1.14 & 0.88 & - & 8.13 \\
TrES-3  & 0.90 & 0.80 & 30 & 1.92 & 0 & 1.31 & 0.04 & 12.8 & 13.78 & 0.80 & - & 10.35 \\
HD41004b  & 0.40 & 0.40 & 50 & 19 & 0.08 & 1.33 & 0.03 & 1400 & 313.59 & 0 & - & 35.35 \\
CoRoT-7  & 0.93 & 0.87 & 23.65 & 0.01 & 0 & 0.85 & 0.04 & 0.3 & 11.24 & 0.68 & - & 12.68 \\
GJ-674  & 0.35 & 0.35 & 34.80 & 0.04 & 0.20 & 4.69 & 0.13 & 0.02 & 285.07 & 0 & - & 35.35 \\
HD189733  & 0.82 & 0.79 & 12 & 1.13 & 0.03 & 2.22 & 0.18 & 3.57 & 7.72 & 0.69 & $-1.4 \pm 1.1$ & 14.14 \\
HAT-P-1  & 1.13 & 1.12 & 26.60 & 0.52 & 0.07 & 4.46 & 0.17 & 0.67 & 5.64 & 0.99 & $3.7 \pm 2.1$ & 9.28 \\
HD149026  & 1.30 & 1.50 & 12.30 & 0.36 & 0.18 & 2.88 & 0.23 & 3.0 & 1.35 & 0.85 & $1.9 \pm 6.1$ & 8.65 \\
TrES-2  & 0.98 & 1 & 25.50 & 1.20 & 0.08 & 2.47 & 0.10 & 5.03 & 7.08 & 0.76 & $-9 \pm 12$ & 9.79 \\
TrES-1  & 0.87 & 0.82 & 32 & 0.61 & 0 & 3.03 & 0.09 & 0.91 & 17.33 & 0.68 & $30 \pm 21$ & 12.83 \\
CoRoT-1  & 0.95 & 1.11 & 11 & 1.03 & 0 & 1.51 & 0.14 & 12.94 & 2.38 & 0.76 & $77 \pm 11$ & 9.38 \\
HAT-P-7  & 1.47 & 1.84 & 19 & 1.80 & 0 & -2.20 & -0.12 & 20.04 & 1.27 & 0.91 & $182.5 \pm 9.4$ & 7.97 \\
WASP-14  & 1.32 & 1.30 & 13.50 & 7.72 & 0.09 & 2.24 & 0.17 & 47.84 & 1.73 & 0.91 & $-33 \pm 7$ & 7.59 \\
\hline
\\
\end{tabular}  }
%\end{center}
{\scriptsize

The masses $M_1$ of the stars are given in Sun mass unity $M_s$, the masses of the exoplanets $M_2$ are given in Jupiter mass unity $M_j$, the radius $R_1$ of the stars is given in Sun radius $R_s$, the spin and orbital periods are in days ($d$), $\eta$ is the RCZ, the molecular kinematic viscosity is in $cm^2\ s^{-1}$ and the typical (inviscid) average growth time $\langle \tau \rangle$ is in thousand years. The molecular kinematic viscosity is calculated considering the molecular kinematic viscosity of the Sun in this zone and assuming a temperature dependence in $T^{-5/2}$, which leads \cite[see e.g.][]{Rieutord} to $\nu=0.001\ (5800/\varsigma)^{5/2}$, using the surface temperature $\varsigma$ of the star in Kelvin units and $\nu$ in $m^2. s^{-1}$.}
  \end{table*}

\subsection{Spin-orbit misalignment: elliptical instability in the star?}

\begin{figure}
  \begin{center}
     \includegraphics[scale=0.5]{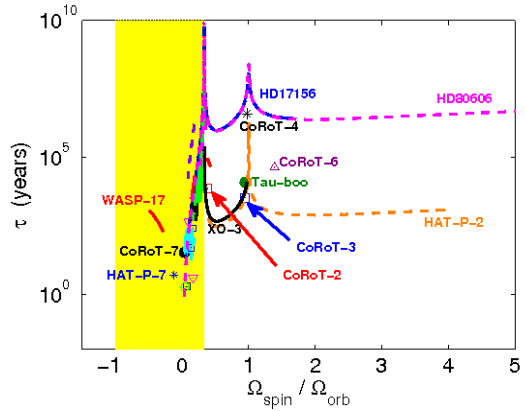}
    \caption{(Colour online) Typical (inviscid) growth time $\tau$ of the elliptical instability in stars, calculated over a whole orbit. Consequently, when the orbital eccentricity of the hot Jupiter is $0$, the star is represented by a point, whereas elliptic orbits lead to lines. The yellow area corresponds to the forbidden zone, where the elliptical instability cannot be excited.}
    \label{fig:obs01}
  \end{center}
\end{figure}
Considering the spin-orbit misalignment of certain hot-Jupiter systems, we can use the spin-orbit angle as a proxy of the presence of the elliptical instability but in the star this time. Indeed, the excitation of the elliptical instability, especially the spinover mode, could give an apparent tilted rotation axis. To test this hypothesis, we proceed in the same manner as for the hot Jupiter: when the star is outside of the forbidden zone, we estimate the typical growth time $\tau=1/\sigma$ of the elliptical instability using the inviscid form of equation (\ref{eq:sigsimpl_OHP}). All the hot-Jupiter system parameters needed for the calculations of this section are given in table \ref{cebronfig3} and results are shown in figure \ref{fig:obs01}. Due to a selection bias (hot Jupiters have orbital periods less than 5-10 days and exoplanets are mainly found around slowly rotating stars with rotational periods more than 20 days, hence $\Omega / \Omega_{orb} \leq 1/3$), 17 stars lie in the forbidden zone:
WASP-17, WASP-10, WASP-19, WASP-18, Kepler-5, TrES-3, HD41004B, CoRoT-7, GJ-674, HD189733, HAT-P-1, HD149026, TrES-2, TrES-1, CoRoT-1, HAT-P-7, WASP-14.
\begin{figure}
  \begin{center}
     \includegraphics[scale=0.5]{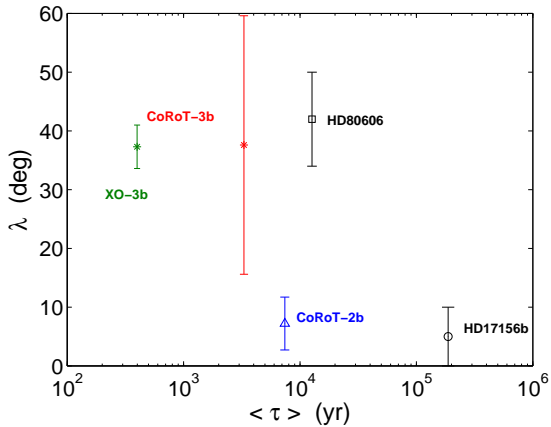}
    \caption{(Colour online) Spin-orbit angle $\lambda$ as a function of the typical inviscid average growth time $\langle \tau \rangle$ of the elliptical instability in the star. Note that $\lambda \neq 0$ is expected for $\tau$ Boo (the absence of transit of the hot Jupiter prevent any measure of $\lambda$).}
    \label{fig:obs2}
  \end{center}
\end{figure}

The stars found to be very unstable are CoRoT-6, $\tau$ Boo, CoRoT-2, CoRoT-3, XO-3 and HD80606 \cite[see also][which shows that the presence of a possible Ekman layer would not drastically change the picture]{cebron2010ohp}. Apart from these ones, HD179949, HD17156, CoRoT-4 could also be unstable (with a lower level of confidence). In figure \ref{fig:obs2}, we test the possibility of a correlation between $\langle\tau\rangle$ and the spin-orbit angle $\lambda$. Here again, although a large dispersion, a general trend is visible and the elliptical instability could be responsible of the misaligned spin-orbit axes of XO-3b, CoRoT-3b, CoRoT-2 and HD 80606b, and a misalignment due to elliptical instability may be predicted for $\tau$ Boo and perhaps CoRoT-6. {The weak value of $\lambda$ for CoRoT-2, which is not really considered as a misaligned system, is expected to change on a timescale of $\langle \tau \rangle \approx 7.4\ \textrm{kyr}$ due to the presence of the elliptical instability. Note also that the absence of transit of the hot Jupiter of $\tau$ Boo does not allow the measurement of the planetary radius or the angle $\lambda$, which is expected to be different from zero due to the presence of the instability.} The spin-orbit misalignment of WASP-17b, CoRoT-1b, HAT-P-7b a,d WASP-14b (all included in our sample but in the forbidden zone) could be explained by an evolution of the system involving elliptical instability occurring before the stellar spin-down.

In this section, we have shown that the hypothesis of the elliptical instability being the main tidal dissipation process in eccentric inflated hot Jupiters and misaligned stars is consistent with the current available data. However, if the stellar rotation axis is tilted by the elliptical instability, leading to a misaligned system, the angular momentum conservation implies some changes for the parameters of the system. In the following section, we confirm that the consequences imposed by the angular momentum conservation do not allow to discard the presence of the instability or inversely that the presence of the instability will not eject the planet to irrealistic very large orbit radius.

\section{Influence on the hot Jupiter orbit of a change in the orientation of the stellar rotation axis} \label{sec:jup_chaud}

In this section, we assume that the elliptical instability is excited, for instance under the form of the spinover mode, which completely modifies the rotation axis orientation of the star. The conservation of the total angular momentum implies a strong coupling between the star and its orbiting planet. In our simulations, a torque can be exerted by the outer ellipsoidal boundary, giving rise to the spinover mode. Whether the gravitational interaction with the planet is sufficient to actually maintain this torque and make the growth of the spinover mode possible still needs to be demonstrated with free surface numerical simulations \cite[][have already shown that the elliptical instability can grow in a Riemann ellipsoid with a free surface, i.e. within a deformable boundary]{ou2007further}. Considering the fact that the growth timescale computed in the last section are comparable to the empirically determined stellar spin tidal evolution timescales, this possibility is not to be excluded.

In this case, as we already saw, the elliptical instability, expected inside certain stars \cite[][]{cebron2010ohp} would be at the origin of the spin-orbit misalignment. Then, we can wonder if the angular momentum conservation of the binary system does not lead to a very large hot-Jupiter orbit, which would be in contradiction with the observations. To tackle this problem, we consider a very simplified model with only a star and one hot Jupiter. Using a subscript $s$ for the star and a subscript $p$ for the planet, we denote by $M$ the mass, by $I$ the inertia momentum, by $\Omega$ the spin rotation rate vector and by $R$ the radius of each body. $a$ is the orbital semi-major axis and $e$ the orbital eccentricity. The evolution of the system is then given \cite[e.g.][]{rieutord2003evolution} by the conservation of the total angular momentum $\mathbf{L}$ of the system
\begin{eqnarray}
\mathbf{L}=\frac{M_s M_p}{M_s + M_p}\ a^2\ \mathbf{\Omega_{orb}}\ \sqrt{1-e^2}+I_s\ \mathbf{ \Omega_s}+I_p\ \mathbf{\Omega_p} \label{eq:momen_anne}
\end{eqnarray}
%and the dissipation of its mechanical energy, $E_m$, given by
%\begin{eqnarray}
%E_m=-\frac{G\ M_s M_p}{2\ a}+\frac{1}{2} I_s\ {\Omega_s}^2+\frac{1}{2} I_p\ {\Omega_p}^2\ ,
%\end{eqnarray}
We neglect the spin angular momentum of the planet in equation (\ref{eq:momen_anne}), which is typically $10^4$ times smaller than the other terms. In this study, we focus on orders of magnitude of the different physical variables. Thus we consider that the elliptical instability is excited on timescales sufficiently short to neglect the induced dissipation. 
Using the third Kepler law $\Omega_{orb}^2\ a^3=G(M_s+M_p)$, where $G$ is the gravitational constant.  the momentum conservation equation (\ref{eq:momen_anne}) then becomes
\begin{eqnarray}
I_s\ \Omega_s\ \cos \theta_1+ \xi\ \sqrt{a}\ \cos \theta_2 = L \label{eq:metto1} \\
I_s\ \Omega_s\ \sin \theta_1+ \xi\ \sqrt{a}\ \sin \theta_2 = 0 \label{eq:metto2}
\end{eqnarray}
where $\theta_1$ and $\theta_2$ are respectively the angles of $\mathbf{L}$ with $\mathbf{ \Omega_s}$ and $\mathbf{\Omega_{orb}}$ and where we have defined
\begin{eqnarray}
\xi= M_s M_p\ \sqrt{\frac{G\ (1-e^2)}{M_s + M_p}}\ .
\end{eqnarray}
\begin{figure}
  \begin{center}
    \begin{tabular}{ccc}
      \subfigure[]{\includegraphics[scale=0.43]{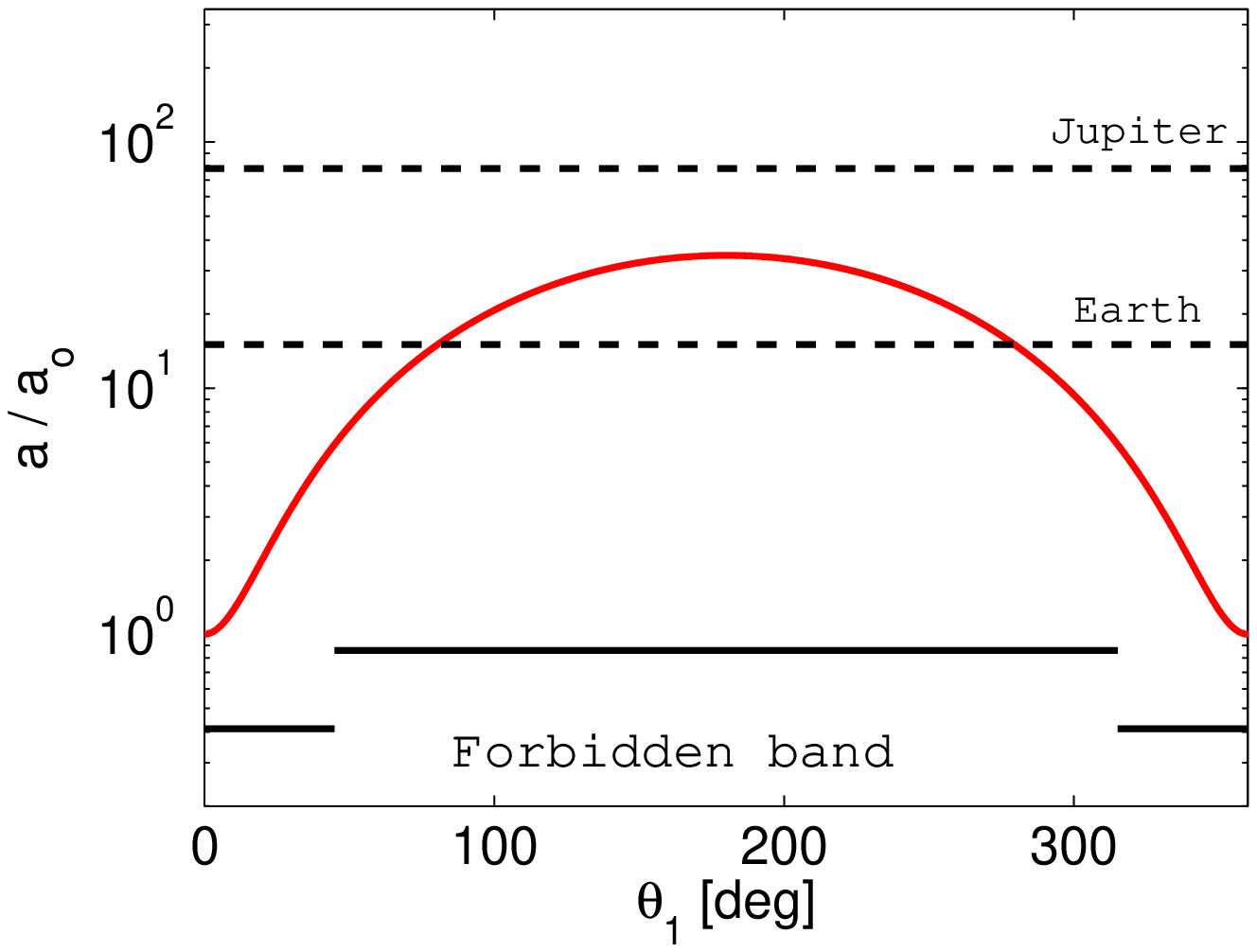}} \\
      \subfigure[]{\includegraphics[scale=0.41]{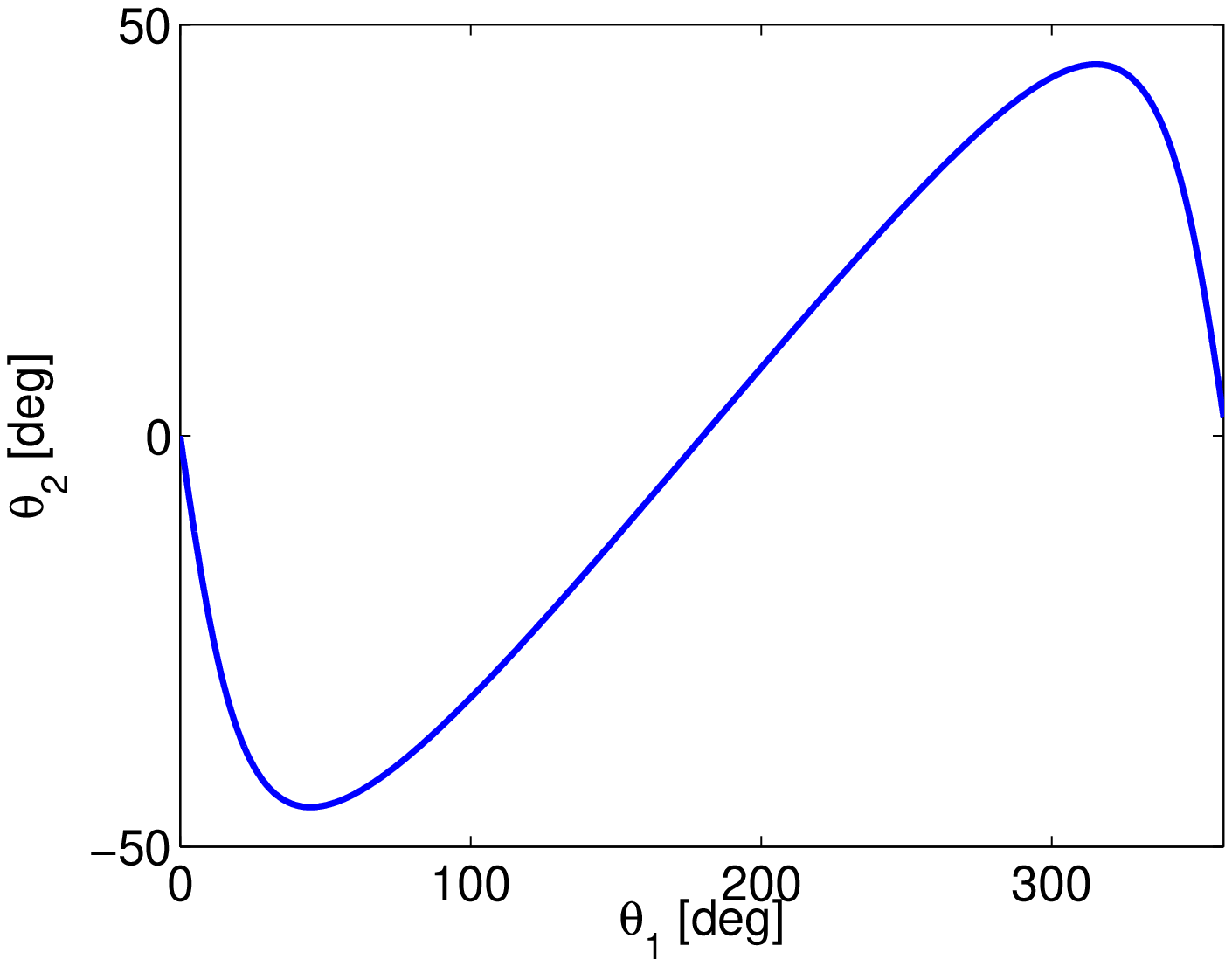}}
     \end{tabular}
\caption{(a) Orbital evolution when $\theta_1$ (deg) is increased ($\theta_1$ is the angle of $\mathbf{L}$ with $\mathbf{ \Omega_s}$). The solid red line represents the orbital semi-major axis (eq. (\ref{eq:demiaxe_annexe})), normalized by the semi-major axis $a_o$ obtained for $\theta_1=0^{\circ}$. The forbidden zone is located between $0$ and the solid black line (first prograde, then retrograde and finally prograde), given by equation (\ref{eq:demiaxe_forbidden}). The dashed lines represent the orbital semi-major axes of the Earth and Jupiter orbits. (b) Evolution of $\theta_2$ when $\theta_1$ (deg) is increased ($\theta_2$ is the angle of $\mathbf{L}$ with $\mathbf{\Omega_{orb}}$).}
    \label{fig:ghjghj}
  \end{center}
\end{figure}
\noindent \noindent Equations (\ref{eq:metto1}) and (\ref{eq:metto2}) lead to the analytical expression of the semi-major axis
\begin{eqnarray}
a=\frac{L^2+I_s^2 \Omega_s^2- 2\  L\ I_s\ \Omega_s\ \cos\ \theta_1}{\xi^2}\ . \label{eq:demiaxe_annexe}
\end{eqnarray}
In figure \ref{fig:ghjghj}a, the evolution of $a$ is represented in a typical case of a star with the mass and the radius of the Sun, spinning in $5$ days (using $I_s=2/5 \cdot M_s R_s^2$, which corresponds to a homogeneous sphere), and a planet with the mass of Jupiter, on a circular orbit of radius $a_o=10^7\ \textrm{km}$ for $\theta_1=0^{\circ}$, which corresponds to a planet approximately $6$ times closer from the star than Mercury from the Sun. In this typical case, the ratio of masses between the planet and the star is about $1000$, and the spin angular momentum of the star represents $71\ \%$ of the total angular momentum of the system. The maximum semi-major axis, when $\theta_1$ is changed, is obtained for $\cos \theta_1=-1$, which gives $a_{max}=(L+I_s\ \Omega_s)^2/\xi^2 \approx 35\ a_o$.

Finally, one can notice that the case of a star which is progressively slowing without changing the orientation of the rotation axis is obtained from equation (\ref{eq:demiaxe_annexe}) with $\cos\ \theta_1=1$:
\begin{eqnarray}
\frac{a}{a_o}=\left[1+\frac{I_s\ \Omega_o}{\xi\ \sqrt{a_o}}\ (1-\cos \alpha) \right]^2\ , \label{eq:demiaxe_annexe2}
\end{eqnarray}
where $\Omega_s(t)=\Omega_o\ \cos \alpha$. This formula systematically leads to smaller orbital semi-major axes than those predicted by formula (\ref{eq:demiaxe_annexe}). {Note however that such spin down usually occurs on timescales for which dissipation and stellar winds are not negligible, and our approximations break down. Indeed stellar winds will remove angular momentum from the star, and both $a$ and $e$ will evolve due to tidal dissipation.}

In a given range of orbital periods, the elliptical instability cannot be excited for any values of $\beta$ and $E$ and $a$ is not modified in this range: this is the so-called forbidden zone, defined by $-1 \leq \Omega_s / \Omega_{orb} \leq 1/3$ \cite[e.g.][]{Cebron_2012}. Using the third Kepler law, this forbidden zone corresponds to an orbital radius $a$
\begin{eqnarray}
0  \leq  a  \leq \left[ \frac{G\ (M_s+M_p)}{\kappa\ {\Omega_s}^2} \right]^{1/3}\ , \label{eq:demiaxe_forbidden}
\end{eqnarray}
where $\kappa=9$ for a prograde orbit et $\kappa=1$ for a retrograde orbit. When $\theta_1$ is modified, the scalar product of equation (\ref{eq:momen_anne}) with  $\mathbf{\Omega_s}$ shows that the orbit is prograde ($\mathbf{\Omega_s} \cdot \mathbf{\Omega_{orb}} >0 $) for $\cos \theta_1 \geq I_s \Omega_s / L$ and retrograde ($\mathbf{\Omega_s} \cdot \mathbf{\Omega_{orb}} <0 $) otherwise, which corresponds to extrema of $\theta_2$, as shown in figure \ref{fig:ghjghj}b. The change of values for the forbidden zone at this transition between a prograde regime and a retrograde regime can also be seen in figure \ref{fig:ghjghj}a.

Finally, if we consider that the elliptical instability modifies the rotation axis of the star, figure \ref{fig:ghjghj} shows that the orbital plane is tilted in the opposite direction in order to maintain the direction of $\mathbf{L}$ , whereas the orbital radius is increased in order to keep the norm of $\mathbf{L}$ constant. However, the largest orbital radius remains in a reasonable range compatible with our solar system dimensions, i.e. between Earth and Jupiter orbital radius (see dotted line of figure \ref{fig:ghjghj}(a). To conclude, note that a change in the orientation of the rotation axis of stars with slower spin rate would lead to smaller orbital radius.

\section{Conclusions}

We have shown that the growth of the elliptical instability remains possible in presence of differential rotation, even with a synchronized latitude, provided that the tidal deformation and/or the rotation rate of the fluid are large enough. Moreover, the amplitude of the instability for a centrally-condensed mass fluid is of the same order of magnitude as the amplitude for an incompressible fluid for a same threshold distance. Possible observational signatures of the elliptical instability have then been considered, using as proxies the radii anomalies for hot Jupiters and the spin-orbit angle for the stars. Having found that the presence of the instability in stars is plausible, we have checked that the angular momentum conservation does not eject the hot Jupiter far from the star in the case where the spin-orbit misalignment would be due to the presence of an elliptical instability in the star.

\section*{Acknowledgements}
  The authors acknowledge M. Rieutord for his numerous valuable suggestions on this study.

For this work, D. C\'ebron was partially supported by the ETH Z\"urich Postdoctoral fellowship Progam as well as by the Marie Curie Actions for People COFUND Program.

% \appendix
%
% \section{Stability analysis for non-synchronized systems} \label{WKB1}

% \bibliographystyle{icarus}
% % Note the spaces between the initials
% \bibliography{convection}

\end{document}